\documentclass[11pt]{article}

\newcommand\eq[1] {(\ref{#1})}

\newcommand{\bfm}[1]{\mbox{\boldmath ${#1}$}}

\newcommand{\beqa}{\begin{eqnarray}}
\newcommand{\eeqa}[1]{\label{#1}\end{eqnarray}}
\newcommand{\bequ}{\begin{equation}}
\newcommand{\eequ}[1]{\label{#1}\end{equation}}
\newcommand{\Grad}{\nabla}

\newcommand{\Ga}{\alpha}

\newcommand{\Gl}{\lambda}

\newcommand{\Gt}{\theta}

\newcommand{\Go}{\omega}

\newcommand{\GD}{\Delta}

\newcommand{\GT}{\Theta}

\newcommand{\GO}{\Omega}

\newcommand{\BGS}{\bfm\Sigma}

\def\Ba{{\bf a}}

\def\Be{{\bf e}}

\def\Bk{{\bf k}}

\def\Bn{{\bf n}}

\def\Bt{{\bf t}}
\def\Bu{{\bf u}}

\def\Bx{{\bf x}}

\def\BC{{\bf C}}

\def\BF{{\bf F}}

\def\BM{{\bf M}}

\def\BR{{\bf R}}

\def\BT{{\bf T}}
\def\BU{{\bf U}}

\newcommand{\beq}{\begin{equation}}
\newcommand{\eeq}{\end{equation}}
\newcommand{\overliner}{\begin{eqnarray}}
\newcommand{\earr}{\end{eqnarray}}
\newcommand{\beqn}{\begin{equation*}}
\newcommand{\eeqn}{\end{equation*}}
\newcommand{\overlinern}{\begin{eqnarray*}}
\newcommand{\earrn}{\end{eqnarray*}}

\newcommand{\fr}{\frac}

\usepackage{color}
\usepackage{graphicx}
\usepackage{subfigure,sidecap,amsmath}

\title{Vortex-type elastic structured media and dynamic shielding}

\author{ {\bf M. Brun$^{(1,2)}$, I.S. Jones$^{(3)}$, A.B. Movchan$^{(4)}$} \\
\\
$^{(1)}${\normalsize {\sl Department of Structural Engineering, University of Cagliari}} \\
{\normalsize {\sl Piazza d'Armi, I-09123 Cagliari, Italy; e-mail: mbrun@unica.it;
}}\\
{\normalsize {\sl web-page: http://people.unica.it/brunmi/ }} \\
$^{(2)}$ {\normalsize {\sl Istituto Officina dei Materiali del CNR (CNR-IOM) Unit\'a SLACS,}} \\
{\normalsize {\sl Cittadella Universitaria, 09042 Monserrato (Ca), Italy}} \\
$^{(3)}${\normalsize {\sl School of Engineering, John Moores University,}} \\
{\normalsize {\sl Liverpool L3 3AF, UK; e-mail: i.s.jones@ljmu.ac.uk}}\\
{\normalsize {\sl web-pages: http://www.ljmu.ac.uk/ENG/72972.htm}}\\
$^{(4)}${\normalsize {\sl Department of Mathematical Sciences, University of Liverpool,}} \\
{\normalsize {\sl Liverpool L69 3BX, U.K.; e-mail: abm@liv.ac.uk}}\\
{\normalsize {\sl web-pages: www.maths.liv.ac.uk/$\sim$abm/}}\\
\\ }
\date{}

\begin{document}

\maketitle

\begin{abstract}

The paper addresses a novel model of metamaterial structure.
A system of spinners has been embedded into a two-dimensional periodic lattice system.
The equations of motion of spinners are used to derive the expression for the chiral
term in the equations describing the dynamics of the lattice.
Dispersion of elastic waves is shown to possess innovative filtering and polarization properties
induced by the vortex-type nature of the structured media.
The related homogenised effective behavior is obtained analytically
and it has been implemented to build a shielding cloak around an obstacle.
Analytical work is accompanied by numerical illustrations.

\end{abstract}

M. Brun et al. Vortex-type elastic structure
\section{Introduction}

Dynamic response of vector elastic lattice systems, interacting with  waves, has been the subject of classical investigations (see, for example, Kunin 1982 and 1983, 
Slepyan, 1981), 
in addition to the more recent publications of Slepyan (2002), Slepyan \& Ayzenverg-Stepanenko (2002) and Colquitt et al. (2011). 
We note substantial differences between scalar problems involving vibration of systems of harmonic springs and vector problems of elasticity, referring to elastic rods and beams, which connect a system of finite solids or point masses. The notion of shear stress and shear strain is absent in scalar lattice systems and there is no such analogue in models of electromagnetism.
The misconception of predictability  of dynamic properties of vector elastic lattices is sometimes based on an intuitive extrapolation of results available for the scalar systems or for problems of electromagnetism  where there is only one wave speed.  A thorough analysis is the only correct way forward and, as illustrated in Colquitt et al. (2011), 
some results for micro-polar elastic lattice structures do not follow from the simpler physical models or intuitive assumptions.

Phononic band gap structures have been analysed in recent years for periodic arrays of voids or finite inclusions in the continuum matrix by Movchan, Nicorovici \& McPhedran (1997), Poulton et al. (2000), Zalipaev et al. (2002), Platts et al. (2002) and Lin \& Huang (2011).
Platts et al. (2003)a and Platts et al. (2003)b studied transmission problems for arrays of elastic structured stacks including a comparative analysis of the filtering properties of elastic waves in doubly periodic media and the transmission properties for the corresponding singly periodic stack structure.
In particular, a review of the Rayleigh Multiple Methods was given by McPedran et al. (2001).
Here, emphasis is on the control of stop bands in the low-frequency region.
Furthermore, a highly cited paper (Milton, Briane \& Willis, 2006) addresses a specially designed elastic coating, with anisotropic inertia properties, which may act as an ``invisibility cloak'', i.e., it routes a wave around an obstacle. Brun, Guenneau \& Movchan (2008) show an alternative in the design  of an invisibility cloak which incorporates a micro-polar composite.

\begin{figure}[ht!]
\centering
{\includegraphics[width=6.cm,angle=0]{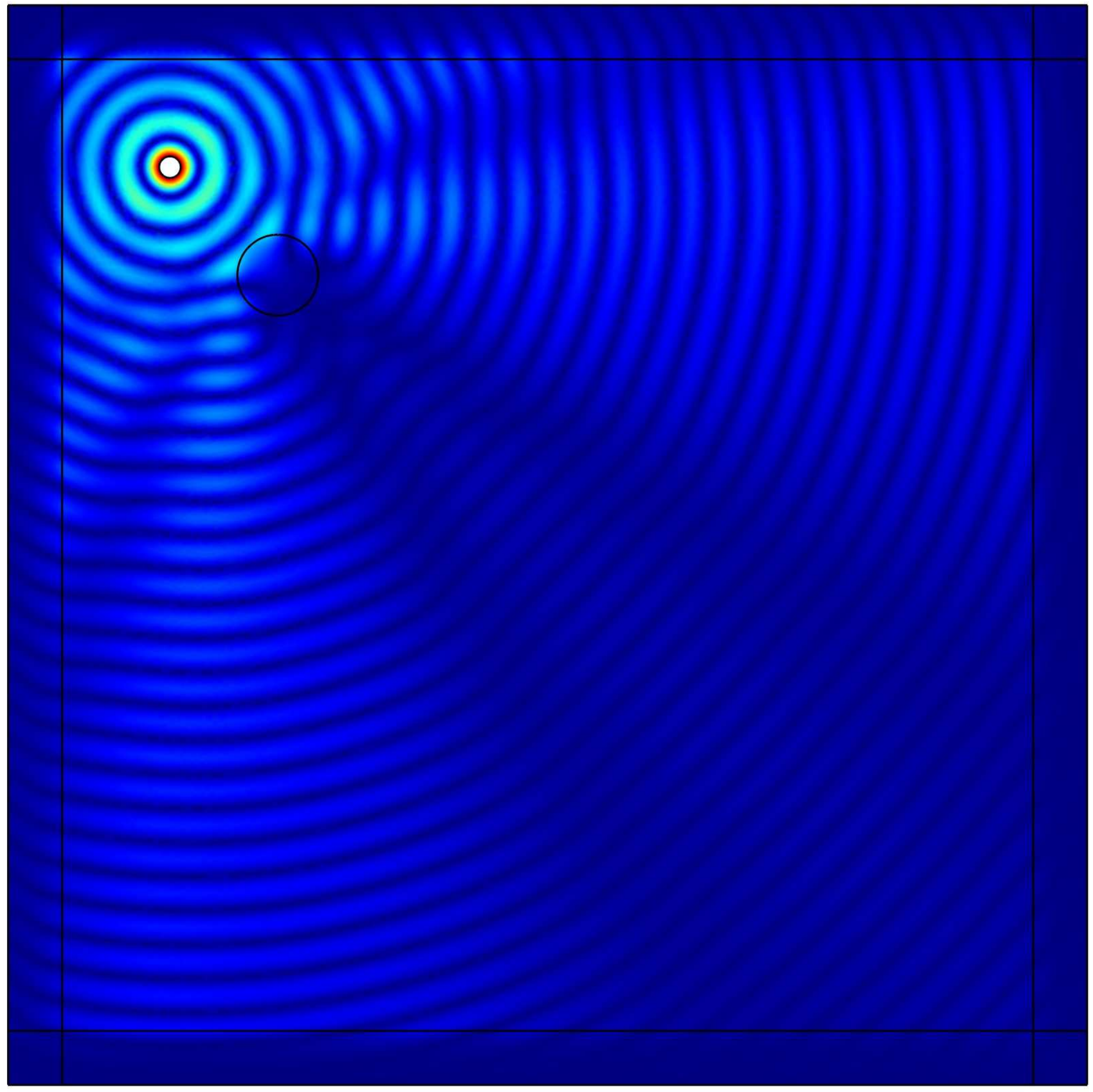}}
{\includegraphics[width=6cm,angle=0]{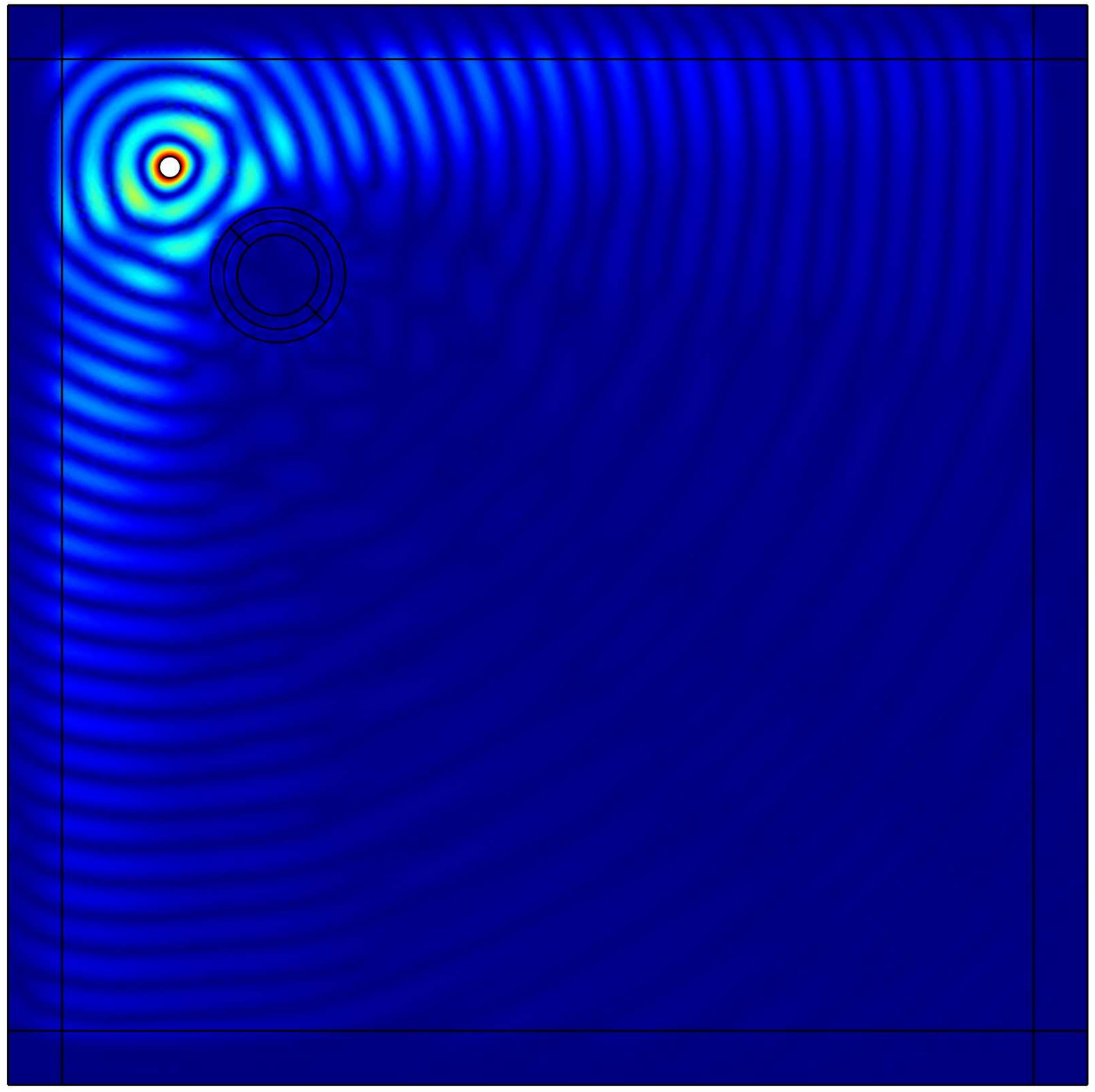}}

{\includegraphics[width=6.cm,angle=0]{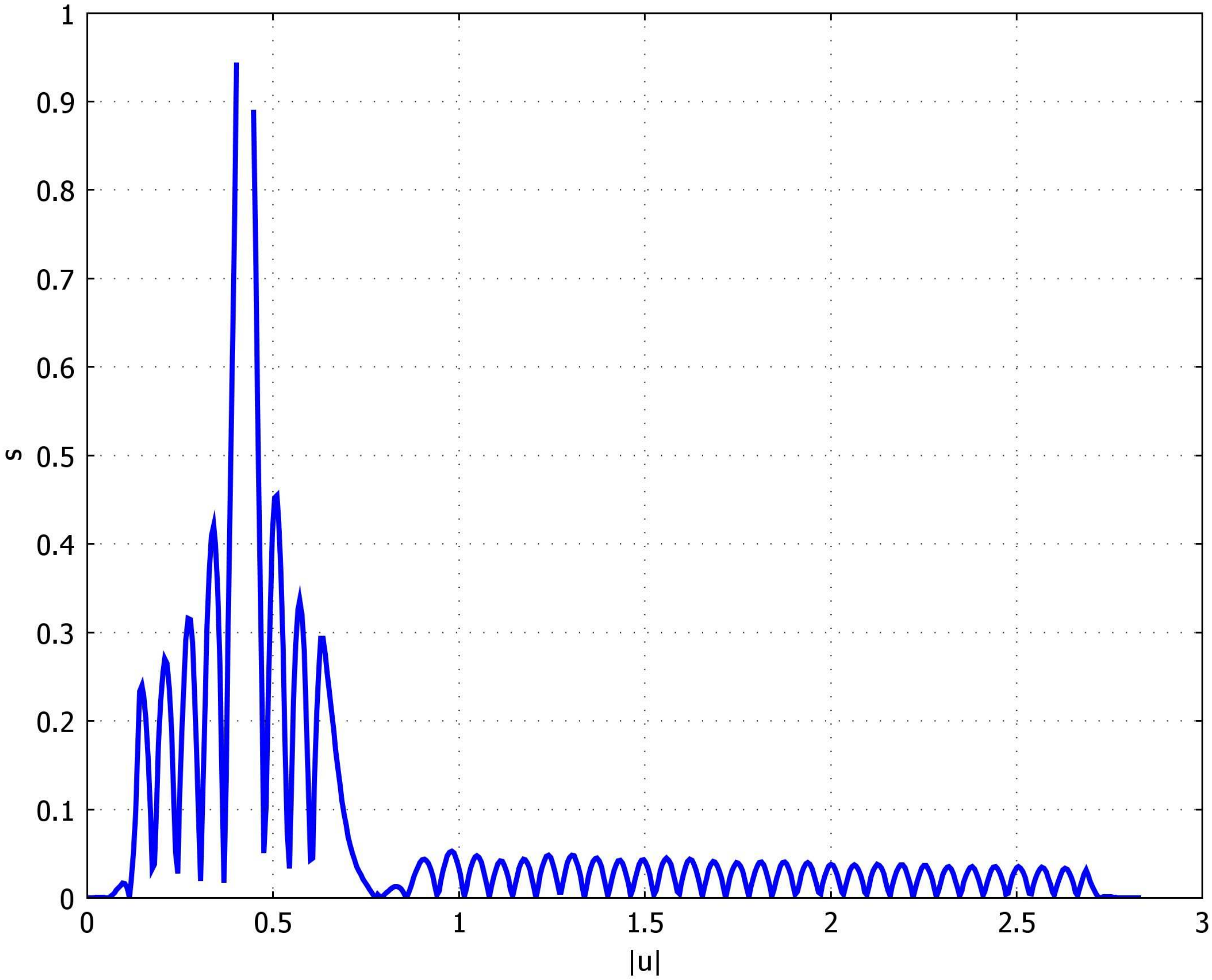}}
{\includegraphics[width=6.cm,angle=0]{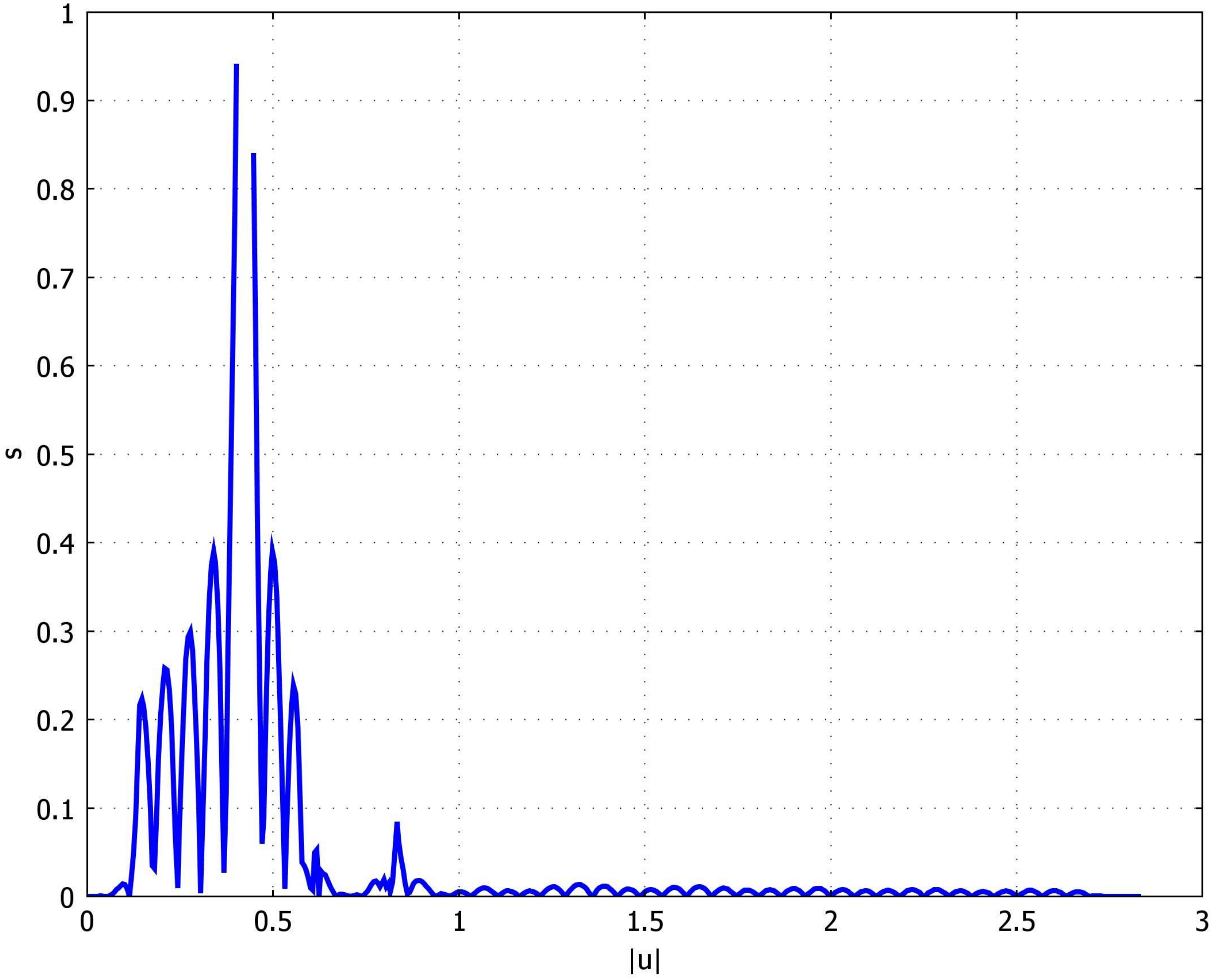}}
\centerline{(a)~~~~~~~~~~~~~~~~~~~~~~~~~~~~~~~~~~~(b) }
\caption{Comparison of the cases  (a) $(\Ga=0.0)$ of the elastic inclusion without coating and (b) the inclusion with a chiral coating $(\Ga=3
0.0)$,  placed in an elastic medium loaded by a time-harmonic point moment.
The chiral coating of large $\Ga$ increases the shaded region created by the inclusion, and suppresses the displacement.
}
\label{force_shear_1}
\end{figure}

It is also feasible to think of a coating around an elastic inclusion, which would enhance (or reduce) the shadow region created as a result of interaction between the inclusion and an incident wave.
Our special interest is in the design of a chiral coating, which could be modelled as a homogenised micro-structure, characterised by a vorticity constant $\Ga$. The numerical simulation in Fig. \ref{force_shear_1} shows as interaction of a radially symmetric shear wave, created by a time-harmonic point moment, with an elastic inclusion; part (a) of the diagram presents the 
displacement amplitude in the plane with a disk-shaped inclusion without a coating; 
part (b) shows the high level of suppression of the scattered  fields when a coated inclusion (with a chiral coating) is introduced in the homogeneous medium.   The graphs in the bottom row of the diagram show the amplitude of the displacement measured  along the main diagonal of the square domain connecting the upper left corner to the bottom right corner of the domain. The displacement behind the inclusion has been  suppressed to a large extent. Needless to say, the applications of this type of design are limitless, with examples including design of earthquake resistant structures in addition to drilling systems in geophysics.

In the present paper, we address propagation of elastic waves, their dispersion properties, and interaction with defects in a two-dimensional structured medium containing vortices. Vortices are created by spinning masses embedded into the lattice system.
The analysis of gyroscopic motion of an individual mass is incorporated into the system of  conservation
of linear and angular momenta
within the lattice systems. Furthermore, elastic Bloch-Floquet waves are considered in  such a system. It is shown that the presence of spinning masses ``stiffens'' the overall system, and it also leads to a special design of chiral media, which possess stop band properties in addition to shielding (re-routing) with respect to  elastic waves. In turn, we consider several types of homogenised chiral systems to design coatings around finite defects in such a way that a dynamic signature of the coated defect may be
enhanced or suppressed compared to the case when the coatings are absent.

\section{A vortex-type lattice system}

A general algorithm for analysing the spectral properties of a periodic lattice was established in Martinsson \& Movchan (2003), and lattice systems with built-in dynamic rotational interactions were analysed in Colquitt et al. (2011).   The new model proposed in the present paper incorporates the  vortex-type of action at the junction regions of the lattice system. We also derive equations containing the vortex terms and explain their physical nature. 

\subsection{Governing equations}
\label{vortex}

\begin{figure}[ht]
\centering
{\includegraphics[width=12cm,angle=0]{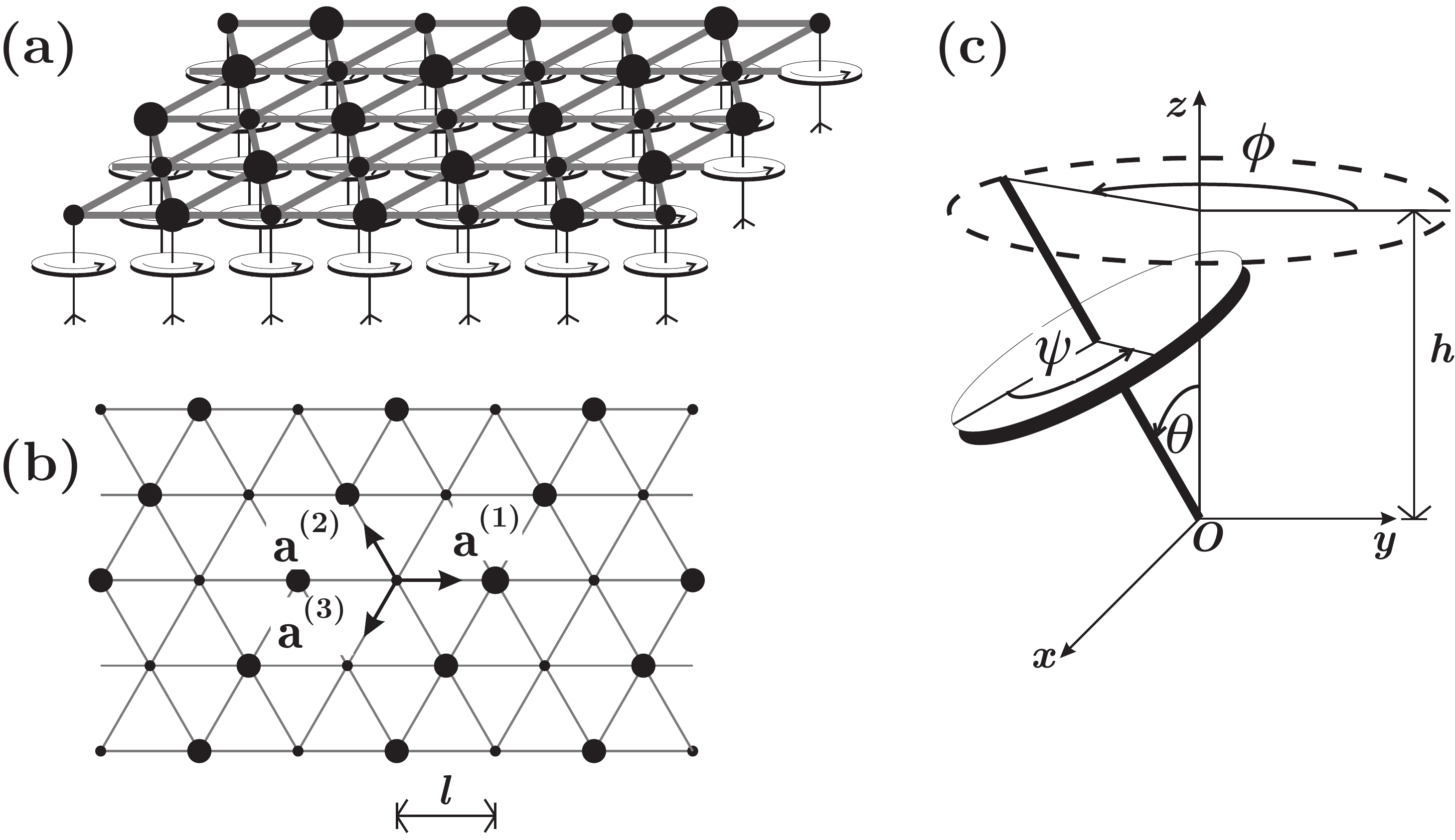}}
\caption{ A triangular lattice linked to the system of spinners is shown in part (a). Part (b) shows the geometry of the triangular bi-atomic lattice. Different masses at junction points are shown as large and small discs. Part (c) shows a spinner and the notations for the coordinate systems; $\psi$, $\phi$ and $\theta$ are the angles of spin, precession and nutation, respectively.
}
\label{fig1}
\end{figure}

We consider a triangular periodic lattice with the elementary cell shown in Fig. \ref{fig1}, which is defined through the basis vectors:
$$
\Bt^{(1)} = (2 l, 0)^T ~~\mbox{and} ~~ \Bt^{(2)}  = (l/2,  l \sqrt{3}/{2})^T.
$$
In this case, we have 
two junctions per elementary cell of the periodic structure, and
it is assumed that the point masses placed at these  junctions are $m_1$ and $m_2$. The stiffnesses of elastic truss-like links, each of length $l$, are assumed to be equal to $c$, and the mass density of the elastic links is assumed to be negligibly small.

We also use the three unit vectors $\Ba^{(j)}, j=1, 2, 3,$ characterising the directions of trusses within the elementary cell:
$$
\Ba^{(1)} = (1, 0)^T, ~ \Ba^{(2)} = (-1/2, \sqrt{3}/2)^T, ~   \Ba^{(3)} = (-1/2, -\sqrt{3}/2)^T.
$$
The multi-index $\Bn = (n_1, n_2)$ with integer entries  is used to characterise the position of the elementary cell within the periodic structure, so that the position vector of the mass $m_\kappa$ is defined as
$$
\Bx^{(\Bn, \kappa)}= \Bx^{(0, \kappa)} + n_1 \Bt^{(1)} +  n_2 \Bt^{(2)},
$$
and
its displacement $\Bu^{(\Bn, \kappa)}e^{i\Go t}$ is time-harmonic, of radian frequency $\Go$.
For convenience, we also use the multi-indices $\Be_1 = (1,0)$ and $\Be_2 =(0,1).$

The design of the lattice system  considered here includes gyroscopic spinners attached to every junction. The axis of each spinner is perpendicular to the plane of the two-dimensional lattice, and its angular velocity can be chosen accordingly by comparison with the radian frequency $\Go$ of the main lattice system. A small  in-plane displacement of a junction leads to a change of orientation of the spinner axis and hence generates a moment creating a ``vortex-type'' effect.

The equations of motion for the masses $m_1$ and $m_2$ within the elementary cell $\Bn$ are
$$
- \fr{m_1 \Go^2}{c} \Bu^{(\Bn, 1)} = \Ba^{(1)} \cdot (\Bu^{(\Bn, 2)} -    \Bu^{(\Bn, 1)} )  \Ba^{(1)} +  (-\Ba^{(1)}) \cdot (\Bu^{(\Bn-\Be_1, 2)} -    \Bu^{(\Bn, 1)} )  (-\Ba^{(1)})
$$
$$
+  \Ba^{(2)} \cdot (\Bu^{(\Bn - \Be_1+ \Be_2, 2)} - \Bu^{(\Bn, 1)}) \Ba^{(2)} +    (-\Ba^{(2)}) \cdot (\Bu^{(\Bn - \Be_2, 2)} - \Bu^{(\Bn, 1)}) (-\Ba^{(2)})
$$
$$
+ \Ba^{(3)} \cdot (\Bu^{(\Bn - \Be_2, 1)} - \Bu^{(\Bn, 1)}) \Ba^{(3)} +  (-\Ba^{(3)}) \cdot (\Bu^{(\Bn + \Be_2, 1)} - \Bu^{(\Bn, 1)}) (-\Ba^{(3)} )
$$
\begin{equation}
{ + \frac{\alpha_1 i \Go^2}{c} \BR \Bu^{(n,1)},}
\label{motion1}
\end{equation}
and
$$
- \fr{m_2 \Go^2}{c} \Bu^{(\Bn, 2)} = \Ba^{(1)} \cdot (\Bu^{(\Bn+\Be_1, 1)} -    \Bu^{(\Bn, 2)} )  \Ba^{(1)} +  (-\Ba^{(1)}) \cdot (\Bu^{(\Bn, 1)} -    \Bu^{(\Bn, 2)} )  (-\Ba^{(1)})
$$
$$
+  \Ba^{(2)} \cdot (\Bu^{(\Bn +\Be_2, 1)} - \Bu^{(\Bn, 2)}) \Ba^{(2)} +    (-\Ba^{(2)}) \cdot (\Bu^{(\Bn +\Be_1- \Be_2, 1)} - \Bu^{(\Bn, 2)}) (-\Ba^{(2)})
$$
$$
+ \Ba^{(3)}  \cdot (\Bu^{(\Bn - \Be_2, 2)} - \Bu^{(\Bn, 2)}) \Ba^{(3)} +  (-\Ba^{(3)}) \cdot (\Bu^{(\Bn + \Be_2, 2)} - \Bu^{(\Bn, 2)}) (-\Ba^{(3)} )
$$
\begin{equation}
 + \frac{\alpha_2 i \Go^2}{c} \BR \Bu^{(n,2)},
\label{motion2}
\end{equation}
where $\BR$ is the rotation matrix
$$
\BR=
\left( {\begin{array}{cc}
 0 & 1  \\
 -1 & 0  \\
 \end{array} } \right)
$$
and $\Ga_1$, $\Ga_2$ are the spinner constants. 

\subsection{Evaluation of the spinner constants}

It is  assumed that each junction within the lattice is connected to a spinner, whose axis is perpendicular to the plane of the lattice, as shown in Fig. \ref{fig1}.
By considering one of the junctions, together with the spinner attached, we evaluate the spinner constants $\Ga_j, j = 1,2,$ introduced in the previous section.

Let $\psi$, $\phi$ and $\theta$ be the angles of spin, precession and nutation with respect to the vertical axis $Oz$, respectively (see, for example, Goldstein et al., 2000, page 210).
An individual spinner and the angles $\psi$, $\phi$ and $\theta$ are shown in Fig. \ref{fig1}c.
Then, assuming  that the gravity force is absent,
the equations of motion of the spinner, for the case of the constant spin rate ($\ddot{\psi} = 0$), are
$$
M_x = I_0 (\ddot{\Gt} - {\dot{\phi}}^2 \sin \Gt \cos \Gt) + I \dot{\phi} \sin \Gt ( \dot{\phi}  \cos \Gt  + \dot{\psi} ),
$$
$$
M_y = \sin \theta \Big(  I_0 (  \ddot{\phi} \sin \Gt + 2 \dot{\phi} \dot{\Gt} \cos \Gt   )   - I \dot{\Gt} ( \dot{\phi} \cos \Gt  + \dot{\psi}  )     \Big),
$$
\begin{equation}
M_z = I (
\ddot{\phi}  \cos \theta - \dot{\phi} \dot{\theta} \sin \Gt ), \label{gyro_motion}
\end{equation}
where $M_x, M_y, M_z$ are the moments about the $x, y$ and $z$ axes,
and $I_0 = I_{xx} = I_{yy}, I = I_{zz}$ are the moments of inertia.

We consider a small amplitude time harmonic motion (due to the motion of the lattice attached to the spinner), resulting in the nutation angle
$$
\Gt(t) = \GT e^{i \Go t}, ~~ |\GT| \ll 1,
$$
and constant spin rate $\dot{\psi} = \GO$.

Assuming that there are no imposed moments about the $x$ and $y$ axes (i.e. $M_x=M_y=0$), and the precession rate is constant ($\ddot{\phi}=0$), we deduce the connection between the spin rate $\GO$ and the radian fequency $\Go$ of the imposed nutation. Namely, the precession rate is given as
\begin{equation}
\dot{\phi}  = \fr{I \GO}{2 I_0 - I}. \label{gyro_1}
\end{equation}
According to the equations of motions we have
\begin{equation}
(I - I_0) {\dot{\phi}}^2
 + I \GO \dot{\phi} - I_0 \Go^2 =0. \label{gyro_2}
 \end{equation}
The compatibility condition for equations \eq{gyro_1} and \eq{gyro_2} has the form
$$
\GO = \pm \Go \fr{2 I_0 - I}{I}.
$$
Direct substitution into \eq{gyro_motion} gives the induced moment about the $z-$ axis
$$
M_z = \mp I \Go \dot{\theta} \theta.
$$
Let $h$ be the characteristic length of the spinner, as shown in
Fig. \ref{fig1}c.
Then the magnitude of the in-plane displacement in the lattice junction is
$$
U = \theta h.
$$
Taking into account that the moment about the $z-$axis is $M_z = F h \Gt$, we deduce
that the ``rotational force'' $F$ is given by
$$
F = \mp I i \fr{\Go^2}{h^2} U.
$$
Following the above derivation, the spinner constants $\Ga_1$ and $\Ga_2$ in equations \eq{motion1} and \eq{motion2} are
$$
\Ga_j = h^{-2} I_z^{(j)}, ~ j = 1,2.
$$

\subsection{Elastic Bloch waves} 

Consider the case when $m_1 = m_2$ and $\alpha_1=\alpha_2$, i.e. the mono-atomic lattice. Then the equations \eq{motion1}, \eq{motion2} are equivalent, subject to a translation  of an elementary cell of the periodic system, which includes only a single junction. This configuration is a good example for observing the influence of the
vortex-type interaction within the lattice on the dispersion properties of the elastic waves.
The amplitudes $U_1, U_2$ of the time-harmonic displacement, of radian frequency $\Go$,   within the periodic vortex-type lattice satisfy the system of equations \eq{motion1}
together with the Bloch Floquet conditions
set within the elementary cell of the periodic system, as follows:
\begin{equation}
\Bu(\Bx + n_1 \Bt^{(1)}/2  + n_2 \Bt^{(2)}) = \Bu(\Bx)  \exp (i \Bk \cdot \BT \Bn),
\end{equation}
where $\Bk$ is the Bloch vector $(k_1,k_2)$, and the matrix $\BT$ has the form
$$
\BT=
\left( {\begin{array}{cc}
 l & l/2  \\
 0 & l \sqrt{3}/2  \\
 \end{array} } \right).
$$

These waves are dispersive, and the corresponding dispersion equation is
\beq
\mbox{det}~ \Big[ \BC(\Bk) - \Go^2 (\BM - \BGS) \Big] = 0,
\eequ{dispersion}
where the mass matrix $\BM=\mbox{diag}[m,m]$ and
$\BGS$ represents the chiral term associated with the presence of spinners attached to the junction points of the lattice system. Assuming that each spinner is characterised by the ``spinner constant'' $\Ga$, 
we represent the matrix $\BGS$ as
\beq
\BGS=
\left(
{\begin{array}{cc}
 0 & -i \Ga  \\
 i \Ga & 0  \\
\end{array} }
\right).
\eequ{vorticity}

Additionally,  in equation \eq{dispersion},  $\BC(\Bk)$ is the stiffness matrix for the monatomic lattice of masses $m$, with spring connectors of stiffness $c$. It is given by
$$
\BC({\bf k})=c\left ( \begin{array}{cccc} 3-2\cos k_1l-\dfrac{(\cos\Phi+\cos\Psi)}{2} & \dfrac{\sqrt{3}(\cos\Psi-\cos\Phi)}{2} \\
\dfrac{\sqrt{3}(\cos\Psi-\cos\Phi)}{2} & 3-\dfrac{3(\cos\Phi+\cos\Psi)}{2}\end{array}\right).
$$
where
$$
\Phi=\frac{k_1l}{2}+\frac {\sqrt{3}}{2}k_2l
\text{\hspace{10pt}and\hspace{10pt}}
\Psi=\frac{k_1l}{2}-\frac {\sqrt{3}}{2}k_2l.
$$

We note that the dispersion equation is bi-quadratic with respect to $\Go$, and
it has the form

\begin{equation}
\Go^4 (m^2 - \Ga^2) - \Go^2     m ~\mbox{tr} \BC + \mbox{det} \BC = 0.
\label{biquad}
\end{equation}

Since $c>0$, both  tr$\BC$ and det$\BC$ are positive for all $k_1$ and $k_2$ in the elementary cell of the reciprocal lattice, except at the origin where they are both zero. Hence there are two important regimes. For $m^2 > \Ga^2$,  there are two dispersion surfaces. As the factor ($m^2 - \Ga^2$) decreases towards a critical point where $m^2 = \Ga^2$, the upper dispersion surface increases without limit and the lower surface decreases. At this critical point, the upper surface is infinite and the lower surface has the degenerate value of
$(m^{-1}~\mbox{det}\BC/\mbox{tr} \BC)^{1/2}$.
For $m^2 < \Ga^2$, the lower dispersion surface only remains. Representative results will now be given for each of the regimes $m^2 > \Ga^2$ (\emph{subcritical}) and $m^2 < \Ga^2$ (\emph{supercritical}).

\subsection{Dispersion properties}

\begin{figure}[ht!]
\centering
 {\includegraphics[width=7cm,angle=0]{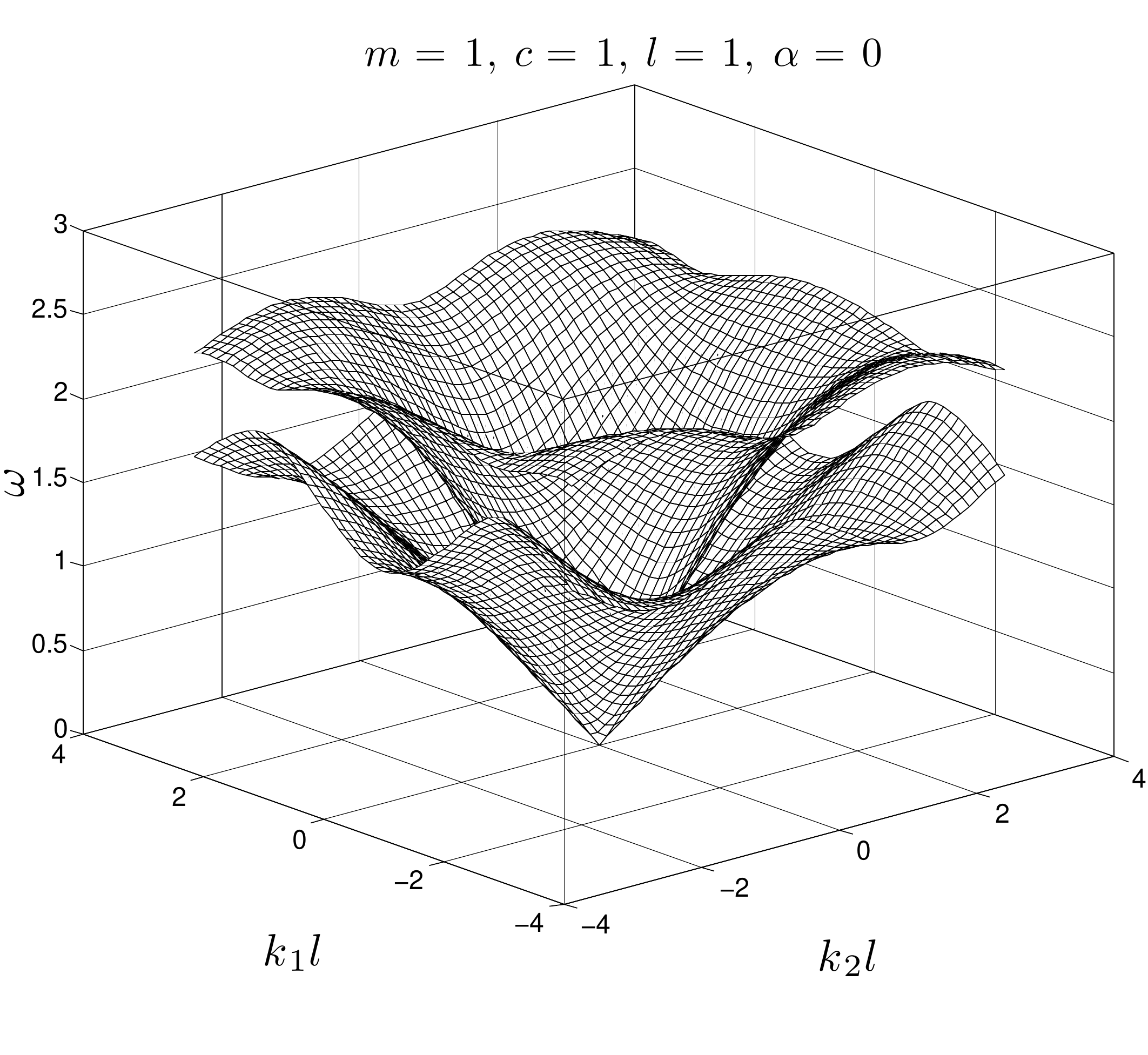}}
\centerline{(a)}
 {\includegraphics[width=9cm,angle=0]{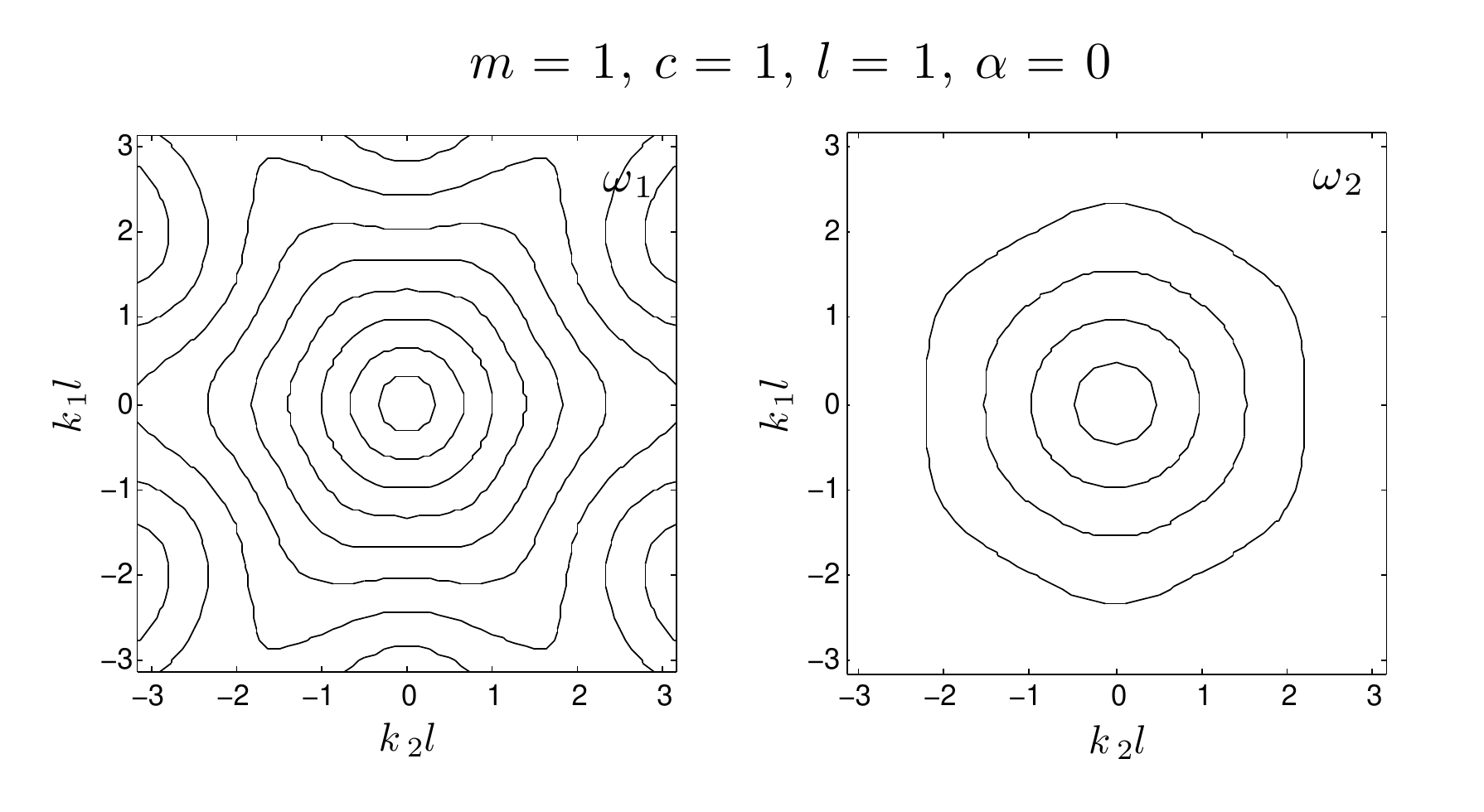}}
 \centerline{(b)}
\caption{
(a)Dispersion surfaces for the monatomic lattice with no spinners.
(b) Contours for the two eigenfrequencies for the monatomic lattice with no spinners}
\label{dispnospin}
\end{figure}

Firstly, Fig. \ref{dispnospin} gives dispersion surfaces and the corresponding two-di\-men\-sio\-nal diagrams
with the slowness contours for the uniform lattice without spinners ($\alpha =0$).
Two conical surfaces are clearly visible in the low frequency range. These surfaces are shown to evolve as
the spinners are brought into the system, now with chiral terms in the equations of motion.

\subsubsection{Monatomic lattice of the vortex-type}

\begin{figure}[ht!]
\centering
{\includegraphics[width=7cm,angle=0]{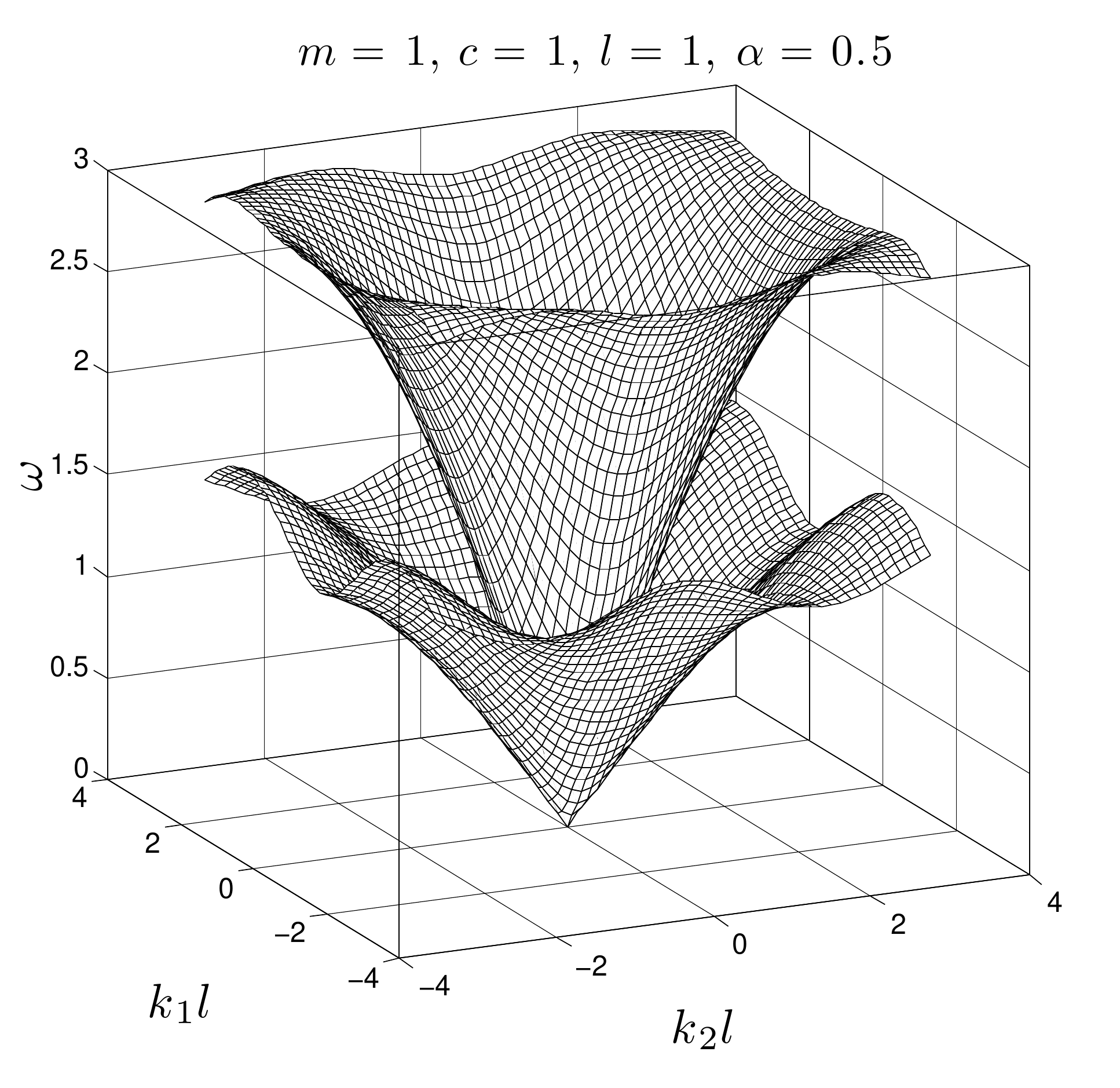}}
\centerline{(a)}
{\includegraphics[width=9cm,angle=0]{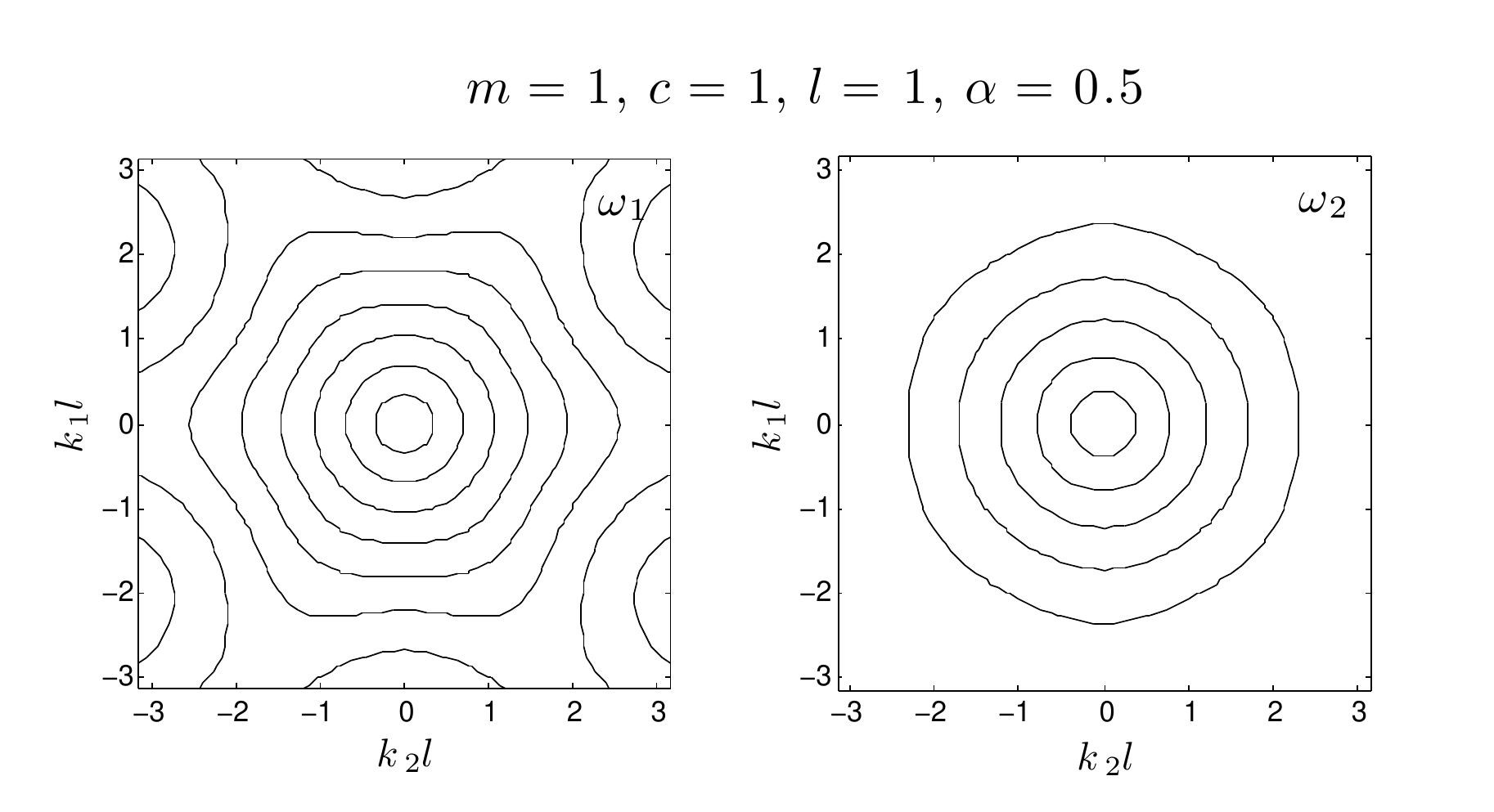}}
 \centerline{(b)}
\caption{
(a) Dispersion surfaces for the monatomic lattice with spinners $\alpha$ = 0.5.
(b) Contours for the two eigenfrequencies for the monatomic lattice with spinners $\alpha$ = 0.5.
}
\label{dispalpha05}
\end{figure}

The diagrams for the vortex-type of monatomic lattice, with a small spinner constant \footnote{
Normalisation has been used throughout the text, so that $c=1$, $m=1$ and the distance between the neighbouring masses is equal to unity.
All other physical quantities have been normalised accordingly, and the physical units are not shown.}
 $\Ga = 0.5$
and mass $m=1$ are given in Fig  \ref{dispalpha05}.
It is apparent that the lower acoustic surface, dominated by shear waves, does not change by much as a result of the vortex interaction. On the contrary, the upper acoustic surface, dominated by pressure waves extends further into a higher range of frequencies and has a pronounced conical shape for a  vortex-type of lattice of small $\Ga$. The latter implies that the homogenisation range for Bloch-Floquet waves corresponding to this dispersion surface is also extended into a wider range of frequencies, as shown in Fig. \ref{dispalpha05}.

\begin{figure}[ht!]
\centering
{\includegraphics[width=7cm,angle=0]{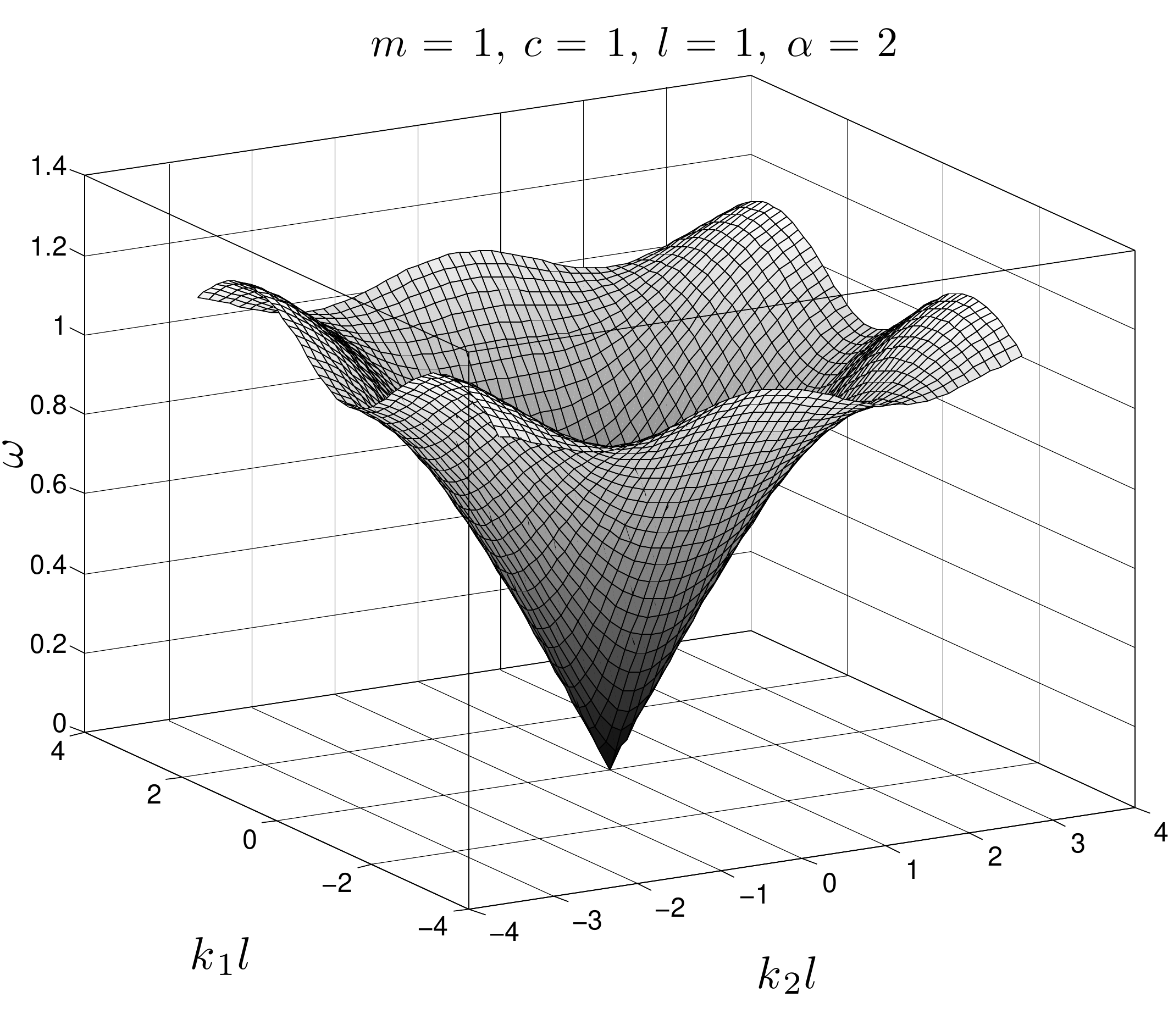}}
\caption{Dispersion surface for the monatomic lattice with spinners $\alpha$ = 2.}
\label{dispalpha2}
\end{figure}

For the other regime, the dispersion diagram illustrating such a situation for the case of $\Ga =2$ and $m=1$ is given in Fig. \ref{dispalpha2}. Only the single dispersion surface is seen for the low frequency range.

\begin{figure}[ht]
\centering
{\includegraphics[width=6.2cm,angle=0]{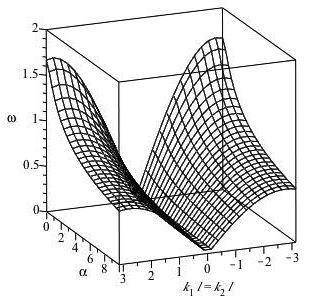}} {\includegraphics[width=6.2cm,angle=0]{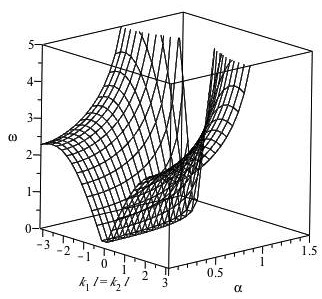}}
\centerline{(a) ~~~~~~~~~~~~~~~~~~~~~~~~~~~~~~~~~~~~ (b)}
\caption{The radian frequency change at the lower dispersion surface (a) and  the upper dispersion surface (b)   with $\alpha$ along the line $k_1=k_2$ in the reciprocal space.
$m=1$, $c=1$ and $l=1$.}
\label{omeg1}
\end{figure}

The influence of the vortex-type of interaction appears to be significant for the dispersion properties of Bloch Floquet waves within the entire admissible frequency range. The illustration is given by Fig. \ref{omeg1}
where we assume $k_1 = k_2 = k$ and plot $\Go$ as a function of the spinner constant $\Ga$ and the Bloch parameter $k$. In this case, the lower dispersion surface is represented by the diagram in Fig. \ref{omeg1}a  and is defined for all real values of 
$\alpha$ whereas the upper  dispersion surface disappears in the supercritical regime when $|\Ga| \geq m$.  The  diagram  characterising the upper dispersion surface and showing $\Go$ as a function of $\Ga$ and $k$ is presented in Fig. \ref{omeg1}b for $0 \leq \Ga <m$  where $m=1$. It is noted that for a fixed value of $k$, the frequency $\Go$ on the upper dispersion surface increases with the increase of the spinner constant $\Ga$ (see Fig \ref{omeg1}b), whereas on the lower dispersion surface the frequency $\Go$ decreases as the spinner constant $\Ga$ increases (see Fig. \ref{omeg1}a).

\subsubsection{The low frequency range}

We examine the behaviour of $\omega({\bf k})$ in the low frequency limit for small values of $|\bf k|$,
and evaluate the effective group velocity, corresponding to a quasi-static response of a homogenised
elastic solid. The solutions to equation (\ref{biquad}) may be expanded for small values of $k_1$
and $k_2$ as follows

\begin{equation}
\omega_1=\sqrt{\frac{3c}{8}\left(\frac{2m-(m^2+3\alpha^2)^{1/2}}{(m^2-\alpha^2)}\right)\big((k_1l)^2+(k_2l)^2\big)},    ~~~~~~~\alpha \neq m,
\label{omeg1a}
\end{equation}

\begin{equation}
\omega_1=\frac{3}{8}\sqrt{\frac{2c}{m}\big((k_1l)^2+(k_2l)^2\big)},    ~~~~~~~\alpha= m,
\label{omeg1meqal}
\end{equation}
 and
$$
\omega_2=\sqrt{\frac{3c}{8}\left(\frac{2m+(m^2+3\alpha^2)^{1/2}}{(m^2-\alpha^2)}\right)\big((k_1l)^2+(k_2l)^2\big)},    ~~~~~~~\alpha < m.
$$

For the case when no spinners are present, these formulae reduce to

$$
\omega_1^0=\sqrt{\frac{3c}{8m}\big((k_1l)^2+(k_2l)^2\big)}    ~~~~\text{and}~~~~~\omega_2^0=\sqrt{3}~\omega_1^0.
$$

These expressions for $\omega_1^0$ and  $\omega_2^0$ are consistent with the special case of equation (3.3) in Colquitt et al. (2011).

In the subcritical regime $m^2 \leq \Ga^2$, for low frequencies, only one dispersion surface exists representing a shear wave.
For a monatomic, harmonic {\it scalar} lattice (with no spinners), the dispersion equation is given in Ayzenberg-Stepanenko  \& Slepyan (2008) as

$$
\omega=\sqrt{\frac{c_s}{m_s}\left(8-4\cos^2\frac{k_1l}{2}-4\cos\frac{k_1l}{2}\cos\frac{{\sqrt 3}k_2l}{2} \right)},
$$
where the subscript $_s$corresponds to the scalar case of out-of-plane shear.
The low frequency approximation for this dispersion equation is

\begin{equation}
\omega=\sqrt{\frac{3c_s}{2m_s}\big((k_1l)^2+(k_2l)^2\big)}.
\label{lowscalar}
\end{equation}

It is interesting to compare equations (\ref{omeg1a}), (\ref{omeg1meqal}) and (\ref{lowscalar}) in the regime $m^2 \leq \Ga^2$. Because of the presence of only one dispersion surface in this regime for the vector lattice of elasticity, then this may be compared with an `equivalent' scalar lattice in which the waves have the same dispersion properties as the shear waves in the vector lattice in the low frequency regime. Comparing equations (\ref{omeg1a}) and (\ref{lowscalar}) and by the appropriate choice

$$
\frac{c_s}{m_s}=\frac{c}{4m}\frac{\sqrt{1+3(\Ga/m)^2}-2}{(\Ga/m)^2-1}
$$
of combination of parameters $c_s$, $c$ and  $m_s$, $m$
for both lattices together with $\Ga$ for the spinners in the vector lattice,  the shear wave in the vector lattice and the wave in the scalar lattice may be constructed so  as to have the same dispersive properties.

\subsubsection{Bi-atomic lattice of the vortex-type}

\begin{figure}[ht!]
\centering
{\includegraphics[width=6.2cm,angle=0]{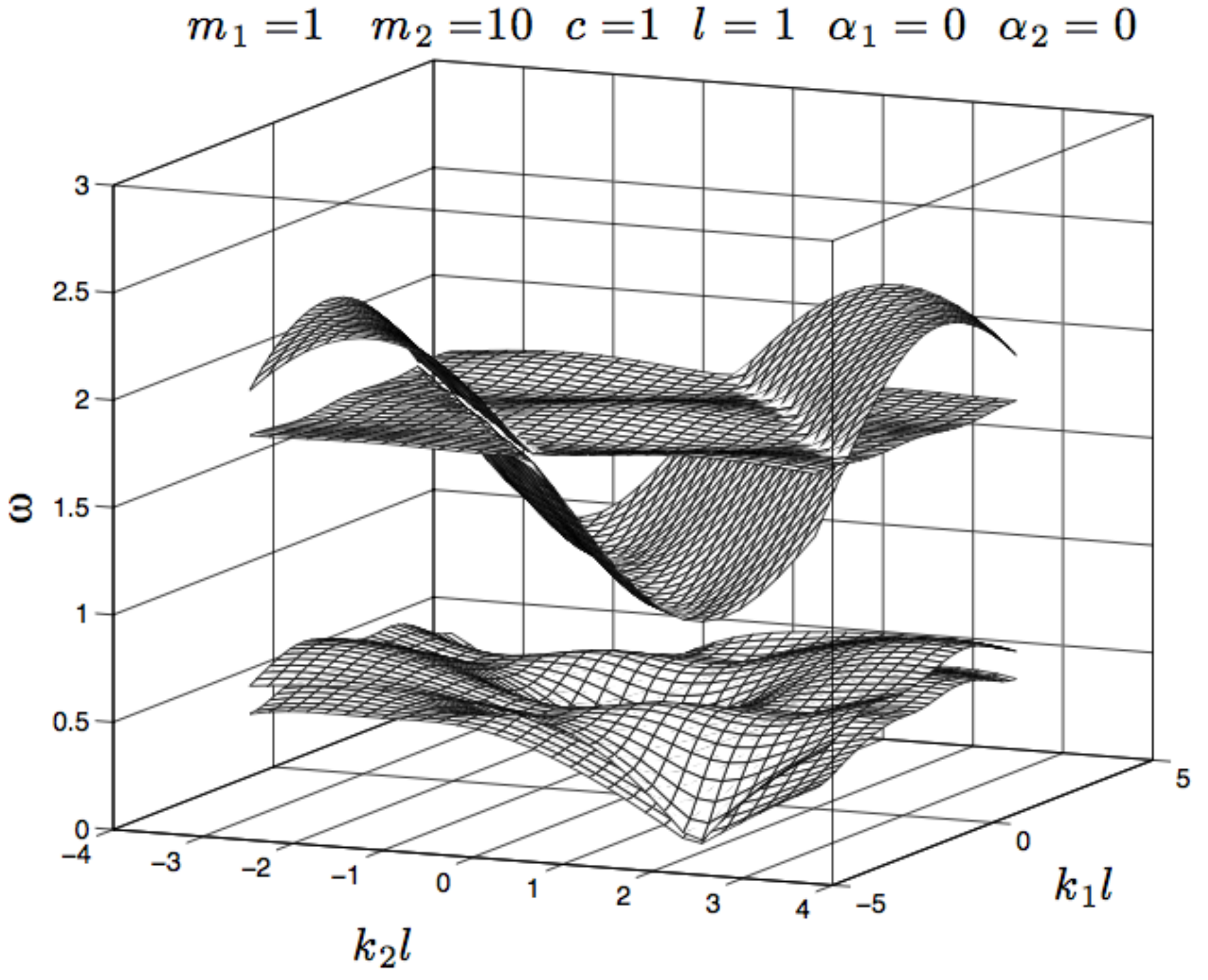}}
{\includegraphics[width=6.25cm,angle=0]{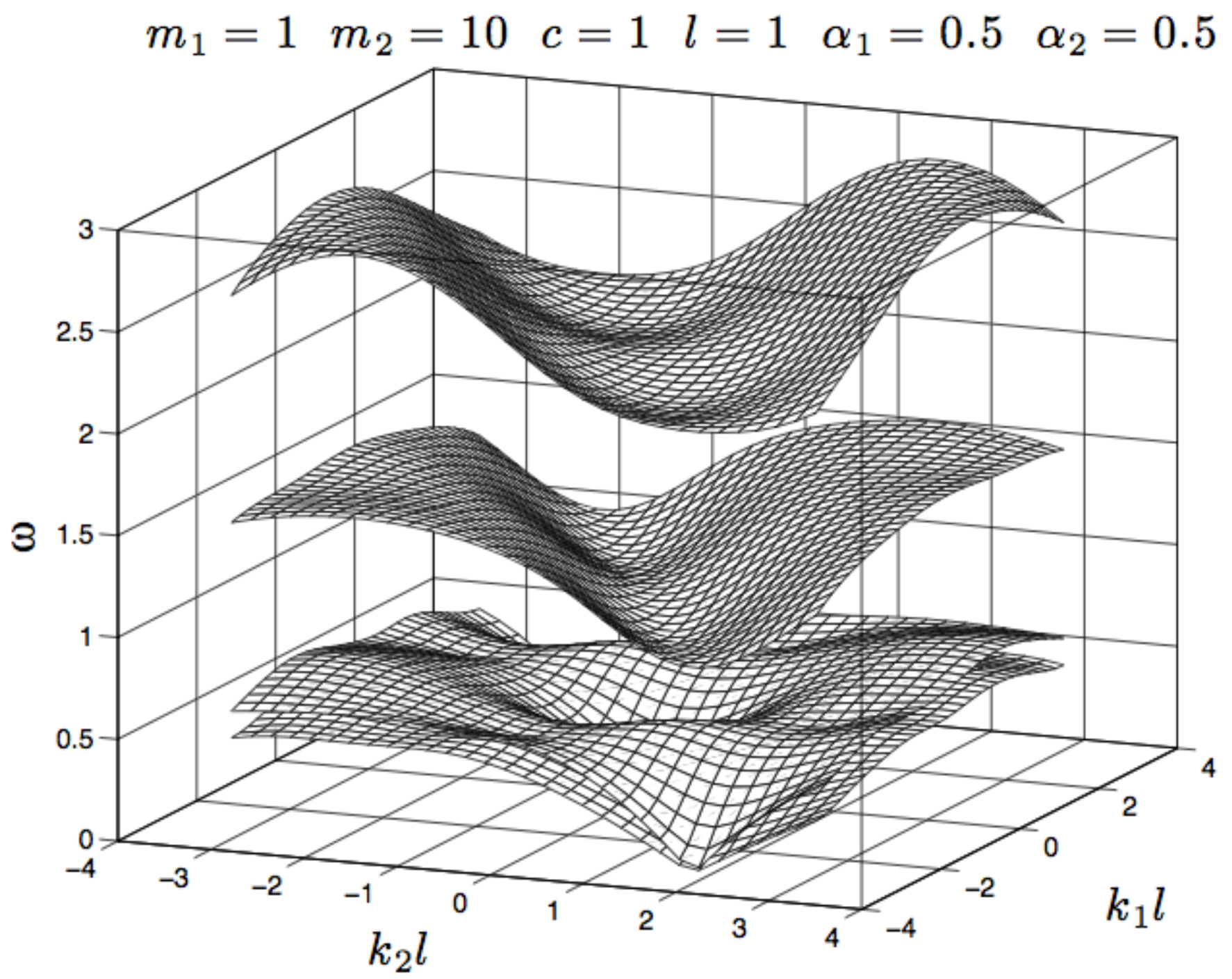}}
\centerline{(a) ~~~~~~~~~~~~~~~~~~~~~~~~~~~~~~~~~~~~~~~~~~ (b) }
\centerline{$\,$}
{\includegraphics[width=5.cm,angle=0]{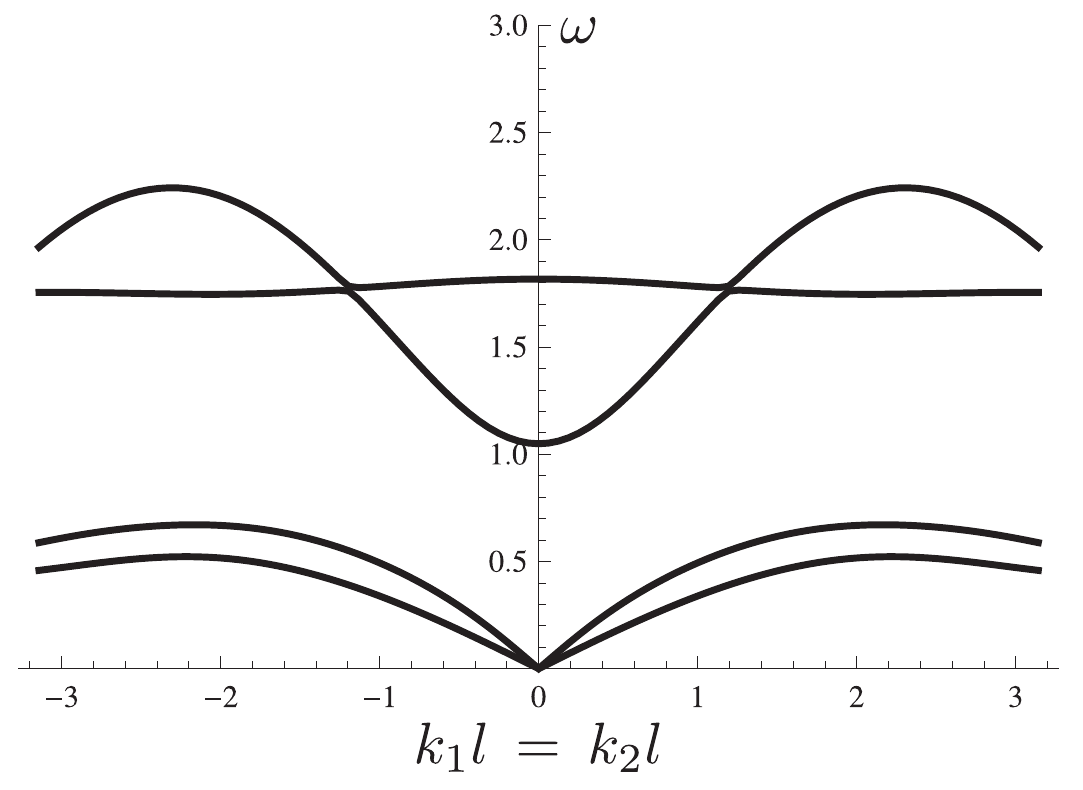}}$\qquad\quad$
{\includegraphics[width=5.cm,angle=0]{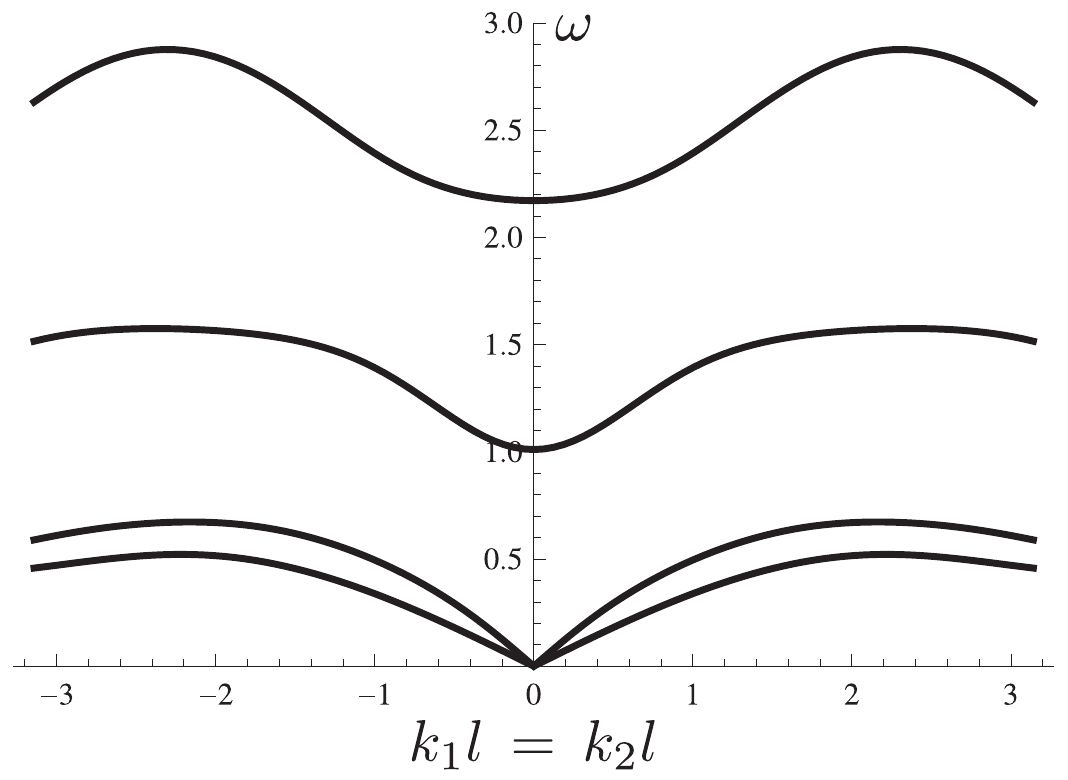}}
\centerline{(c) ~~~~~~~~~~~~~~~~~~~~~~~~~~~~~~~~~~~~~~~~~~ (d) }\\
\caption{
Dispersion surfaces for the biatomic lattice (a) with contrasting masses and no spinners, and (b) with contrasting masses and the same spinner throughout the lattice. (c) and (d) are the cross section of the surfaces given in (a) and (b), respectively, for $k_1\,l=k_2\,l$.}
\label{biadispnospin}
\end{figure}

Now we show the computations for Bloch-Floquet waves for the biatomic triangular lattice described in Section \ref{vortex}.
The dispersion equation is given by relation (\ref{dispersion}), where
$$
\BM=\mbox{diag}[m_1,m_1,m_2,m_2]
$$
and
$$
{\bf \Sigma}({\alpha})= \left( \begin{array}{cccc}
0 & -i\,\alpha_1 & 0 & 0 \\
i\,\alpha_1 & 0 & 0 & 0 \\
0 & 0  & 0 & -i\,\alpha_2 \\
0 & 0  & i\,\alpha_2 & 0
\end{array}
\right).
$$
Additionally, the stiffness matrix can be expressed in the form (see Martinsson \& Movchan 2003)
$$
\BC({\bf k})\!=c\left( \begin{array}{cc}
\BC_{11}({\bf k}) & \BC_{12}({\bf k}) \\
\BC_{21}({\bf k}) & \BC_{22}({\bf k})
\end{array}
\right),
$$
with the $2\times 2$ matrices
$$
\begin{array}{c}
\BC_{11}({\bf k})\!=\BC_{22}({\bf k})=\left( \begin{array}{cc}
3-\frac{1}{2}\cos\Phi & -\frac{\sqrt{3}}{2}\cos\Phi \\[0.08 in]
-\frac{\sqrt{3}}{2}\cos\Phi & 3-\frac{3}{2}\cos\Phi
\end{array}
\right),\\[0.15 in]
\BC_{21}({\bf k})\!=\overline{\BC_{12}({\bf k})}=\exp[i(\Phi+\Psi)]\left( \begin{array}{cc}
-2\cos(\Phi+\Psi)-\frac{1}{2}\cos\Psi & \frac{\sqrt{3}}{2}\cos\Psi   \\[0.08 in]
\frac{\sqrt{3}}{2}\cos\Psi & -\frac{3}{2}\cos\Psi
\end{array}
\right).
\end{array}
$$

For such a lattice with no spinners, the dispersion surfaces for Bloch-Floquet waves are shown in Fig \ref{biadispnospin}a
(a section of the surfaces for $k_1=k_2$ is added in Fig. \ref{biadispnospin}c).
In contrast, for the vortex-type of lattice the corresponding results are given in Fig \ref{biadispnospin}b
(with a section for $k_1=k_2$ in Fig. \ref{biadispnospin}d).

These computations show that the vortex-type of interaction within the lattice leads to
formation of additional total band gaps, as shown in Fig. \ref{biadispnospin}b.
It is noted that the change in the behaviour of the lowest two dispersion surfaces through the introduction of spinners is small,
whereas additional band gaps are introduced in the upper two dispersion surfaces.

\begin{figure}[ht!]
\centering
{\includegraphics[width=6.1cm,angle=0]{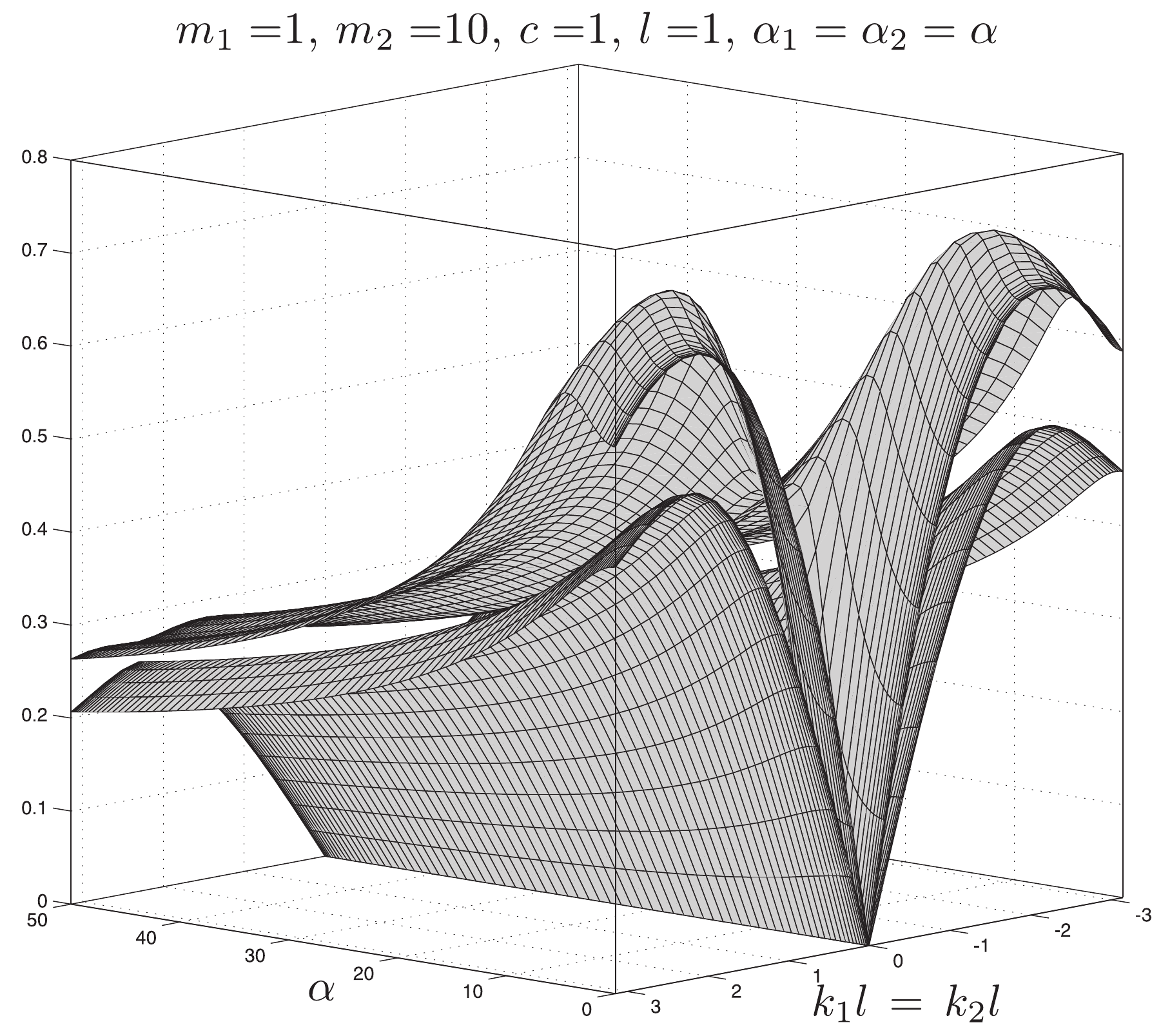}}
{\includegraphics[width=6.1cm,angle=0]{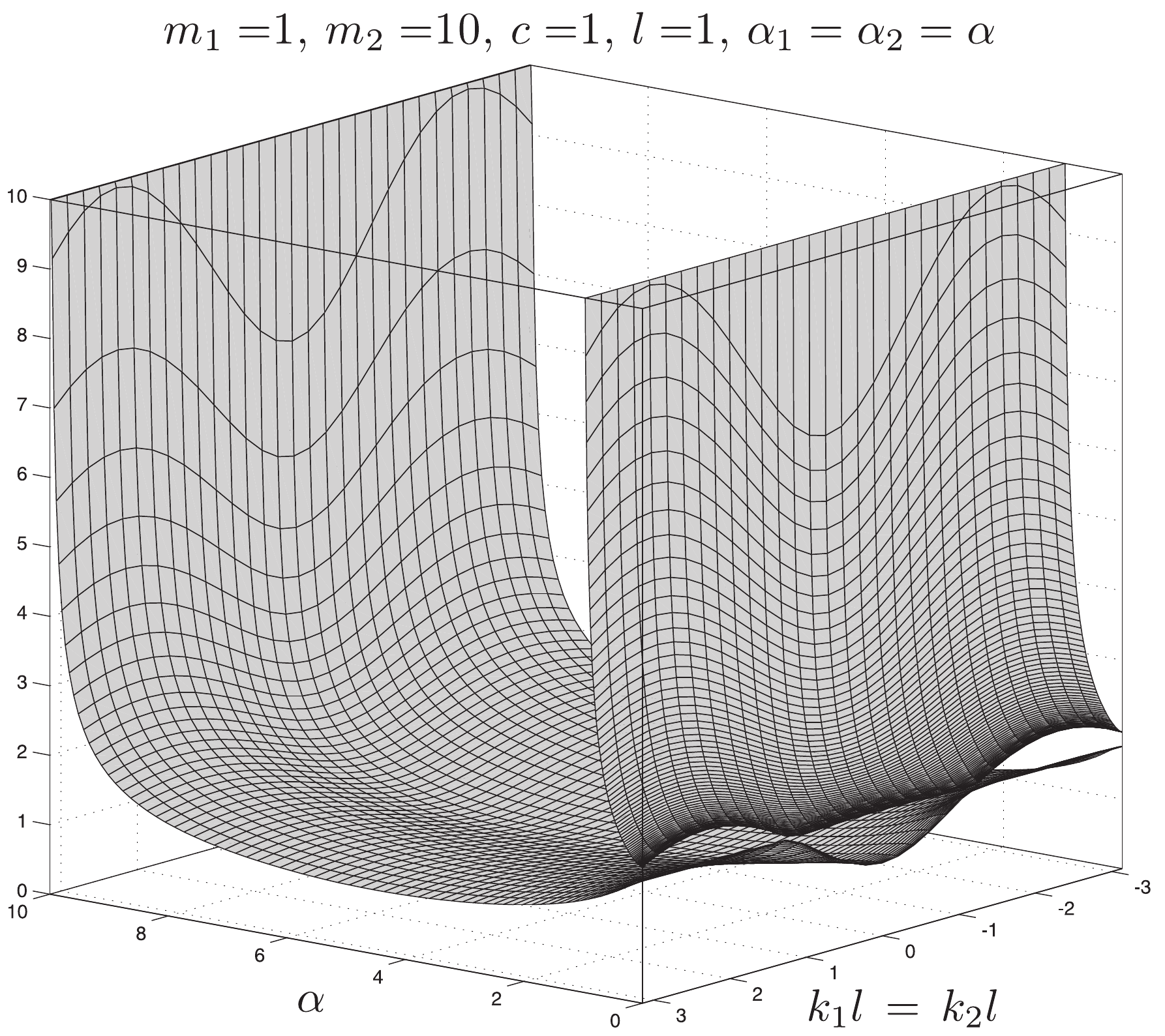}}
\centerline{(a) ~~~~~~~~~~~~~~~~~~~~~~~~~~~~~~~~~~~~~~~~~~ (b) }
\caption{The radian frequency change at the two lower dispersion surfaces (a) and the two upper dispersion surfaces (b) with $\alpha$ along the line $k_1=k_2$ in the reciprocal space.
$m_1=1$, $m_2=10$, $c=1$ and $l=1$.}
\label{biadispalpha}
\end{figure}

The influence of the spinner constant $\alpha$ on the dispersion properties of Bloch-Floquet waves is
illustrated in Fig. \ref{biadispalpha} where $\omega$ is represented as a function of $\alpha$ and $kl=k_1l=k_2l$.
The two lower shear dispersion surfaces, represented in Fig. \ref{biadispalpha}a, are defined for all real values of $\alpha$, whereas the two upper
pressure dispersion surfaces, depicted in Fig. \ref{biadispalpha}b, disappear when critical regimes determined by the spinner constant $\alpha$ are reached.
In particular, in the {\it subcritical regime} ($\alpha\leq 1$) two pressure waves propagate while one disappears in the {\it intercritical regime} ($1<\alpha\leq 10$) and no pressure waves propagate in the {\it supercritical regime} ($\alpha>10$).
Also, for the lower dispersion surfaces, at sufficiently large $\alpha$ the frequency $\omega$ decreases as the
spinner constant $\alpha$ increases (see Fig. \ref{biadispalpha}a).

\section{Homogenisation approximation. Dynamic shielding}

Here we discuss the dynamic response and shielding properties of a  homogenised elastic solid, whose equations of motion contain the vortex-type term.

As illustrated in the previous section the long-wave approximation for the triangular lattice, incorporating the periodic system of spinners, corresponds to an isotropic elastic material with the Lam\'{e} constants $\Gl = \mu$, mass density $\rho$  and an additional term representing the chiral properties of the medium. The amplitude vector $\BU$ of the time-harmonic displacement satisfies the equations of motion as follows
$$
\mu \Big( \GD \BU  + 2 \Grad \Grad \cdot \BU\Big) + \Go^2
 \BGS \BU + \rho \Go^2 \BU  + \BF = 0,
$$
where $\mu$ is the shear modulus of the homogenised solid, $\BGS$ is the vorticity matrix defined in \eq{vorticity},
In the computations presented here, we normalise the mass density and the shear modulus of the chiral medium,
so that $\rho =1.0, \mu =1.0$, and distinguish between  (a) the {\it subcritical} case of $|\Ga| < 1$ and (b) the
{\it supercritical} case of $|\Ga| \geq 1,$ where $\alpha$ is the spinner constant from \eq{vorticity}.
Such a classification is linked to the dispersion equation \eq{biquad}.
In the regime (a), there are two conical surfaces for small values of $\Go$ in the dispersion diagram (see Fig.
\ref{dispalpha05}). On the other hand, in the regime (b)  there is just one acoustic dispersion surface, which has a conical shape for small values of $\Go$.

The chiral medium is used in the coating for an elastic inclusion, which interacts with an incident wave created by point sources of different types.

\begin{figure}[ht!]
\centering
{\includegraphics[width=5.5cm,angle=0]{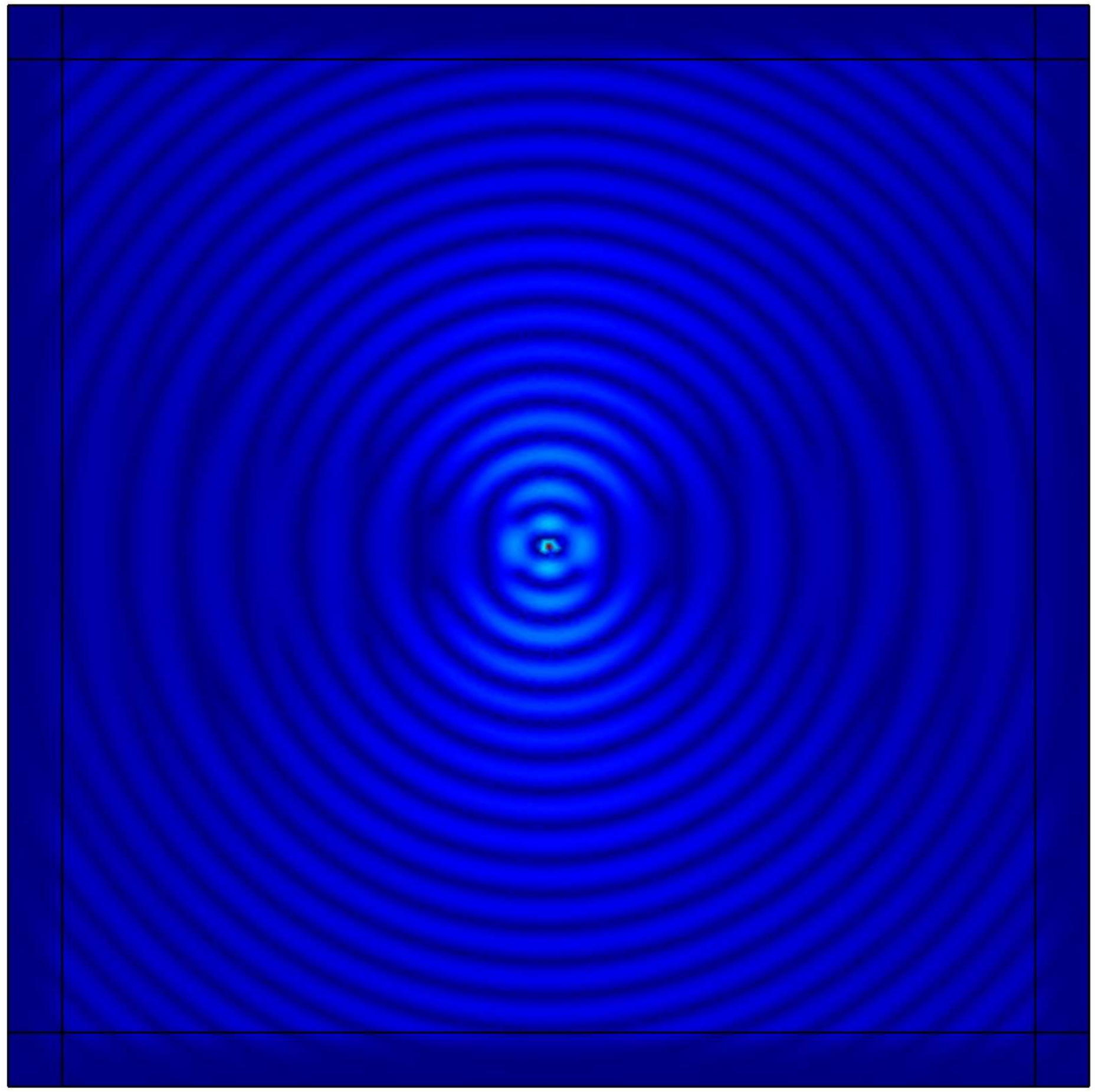}}   {\includegraphics[width=5.5cm,angle=0]{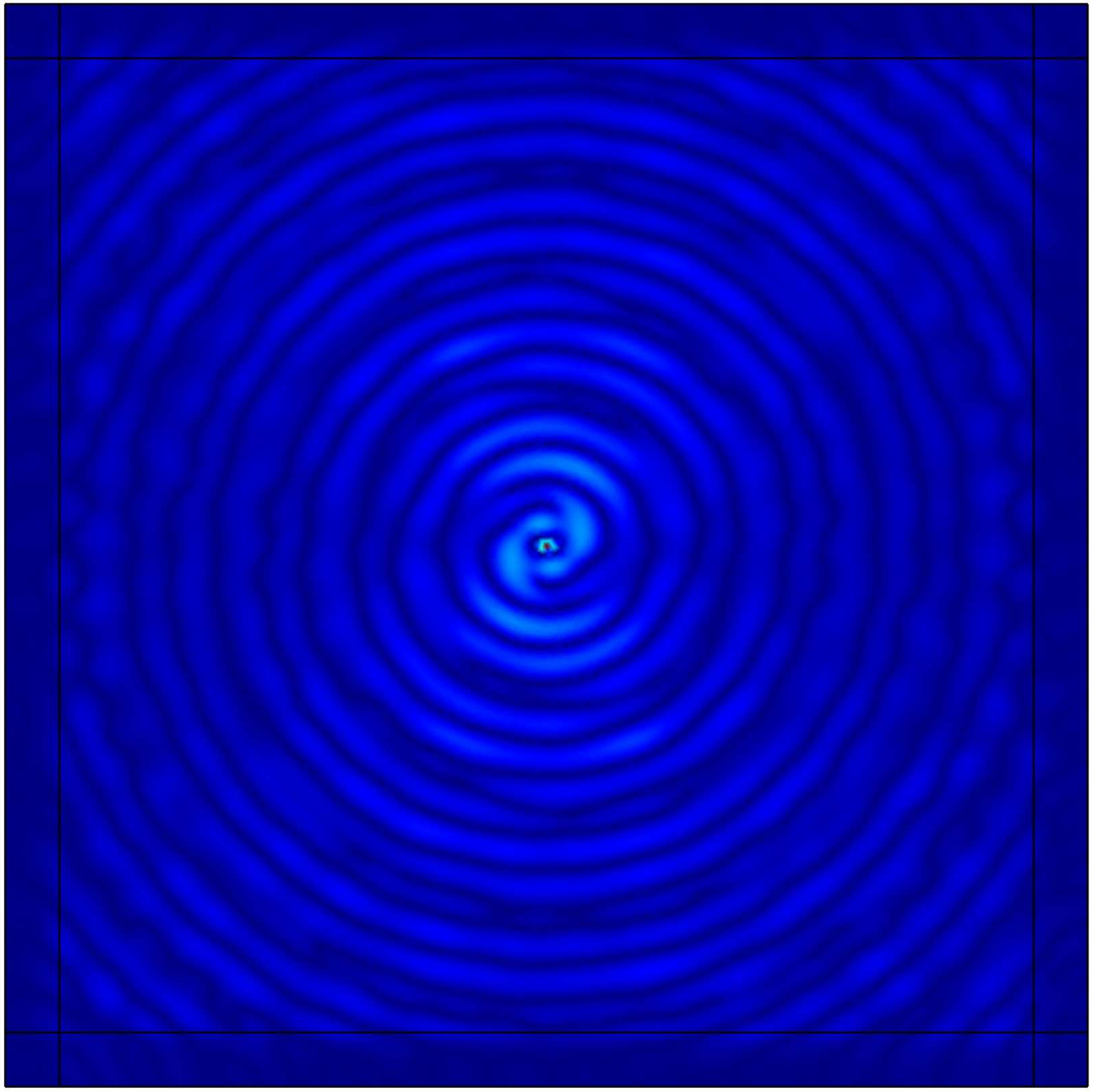}}
\centerline{(a) ~~~~~~~~~~~~~~~~~~~~~~~~~~~~~~~~~ (b) }
 {\includegraphics[width=5.5cm,angle=0]{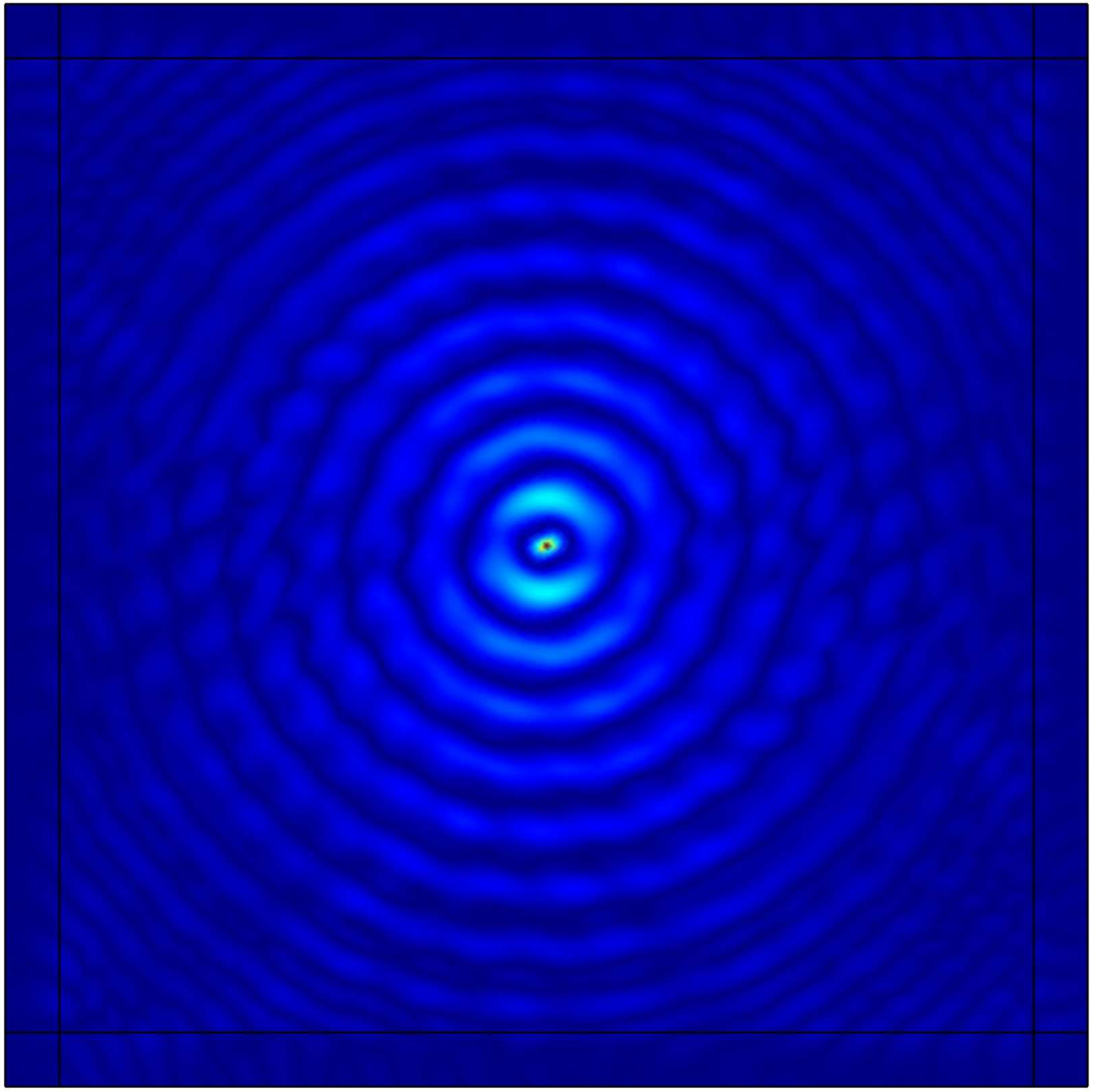}}    {\includegraphics[width=5.5cm,angle=0]{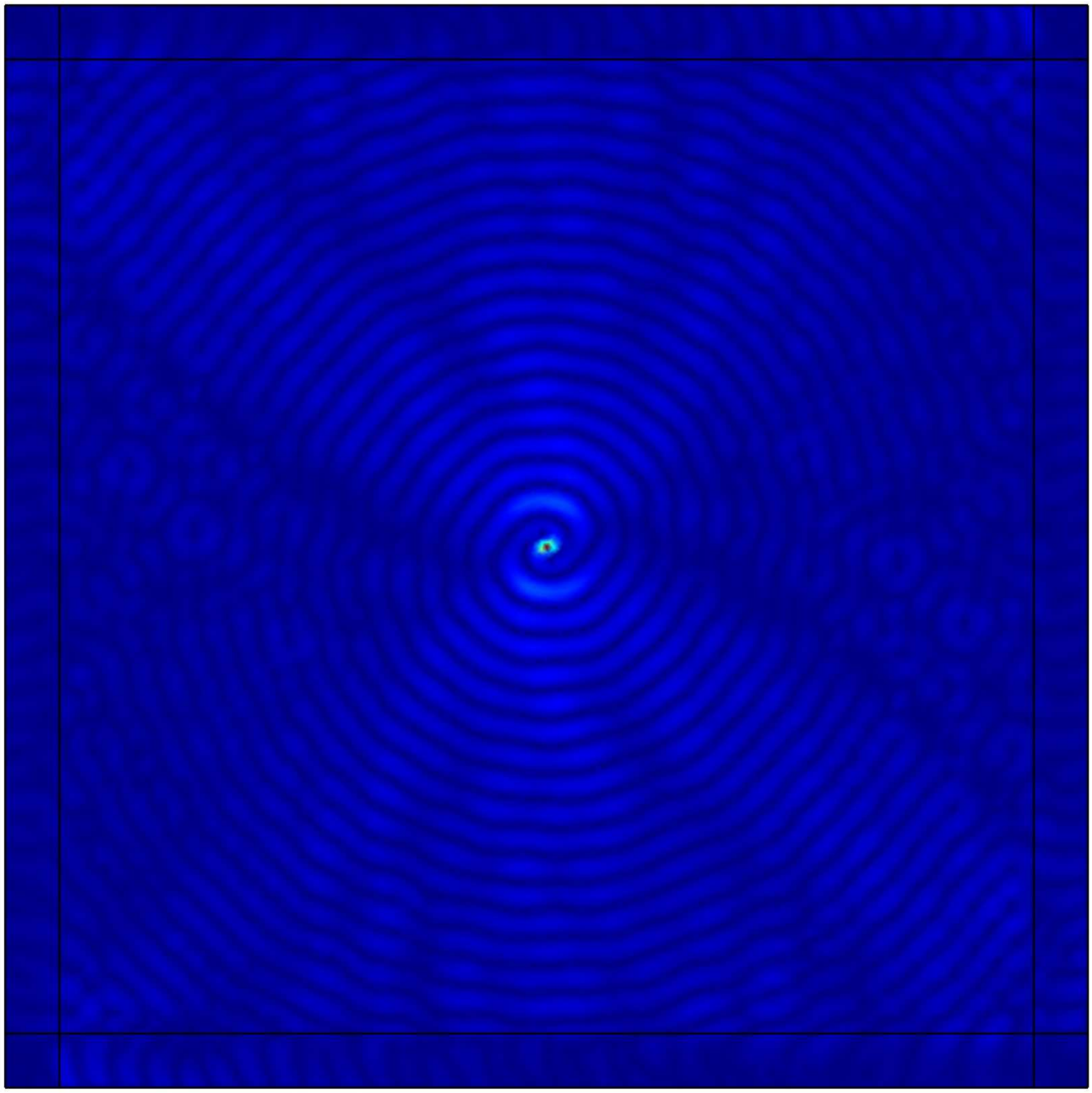}}
  \centerline{(c) ~~~~~~~~~~~~~~~~~~~~~~~~~~~~~~~~~ (d) }
\caption{
 Displacement amplitude produced by  the unit force  applied in the $x_1-$direction
 in the homogenised chiral medium: (a) $\alpha$ = 0.0  (no spinners), (b)  $\alpha$ = 0.6, (c)  $\alpha$ = 1.0, (d)  $\alpha$ = 1.4.   }
\label{force1}
\end{figure}

 \subsection{Point force in the chiral solid}

Firstly, we show the results of the computational experiment produced for a homogenised chiral solid loaded by a time-harmonic horizontal
point force of the radian frequency
$\omega = 50$.

Fig. \ref{force1} shows the change in the wave pattern for different values of the spinner constant $\alpha$.
In particular, Fig. \ref{force1}a has the result of the computation for the non-chiral medium  ($\alpha=0.0$),
whereas Fig. \ref{force1}b includes the modified plot corresponding to a subcritical value of the spinner
constant ($\alpha=0.6$) demonstrating a vortex around the point force.
Domination of pressure waves, with a relatively large wave-length is clearly visible in the horizontal direction
in Fig. \ref{force1}a compared to Fig. \ref{force1}b.

Fig. \ref{force1}c shows the critical case, with the spinner constant $\alpha=1.0$, and the presence of
spinners leads to a nearly isotropic map; the preferential directions associated with the point force are less pronounced.

Furthermore, Fig. \ref{force1}d includes the displacement amplitude for another supercritical case with the
larger value $\alpha=1.4$, and the dominance of shear waves, with the relatively small wave length is clearly visible.

The computations suggest that the directional preference is substantially reduced in the presence of  `spinners'
in the chiral medium in the supercritical regime of $|\Ga| \geq 1.$

\begin{figure}[ht!]
\centering
{\includegraphics[width=10cm,angle=0]{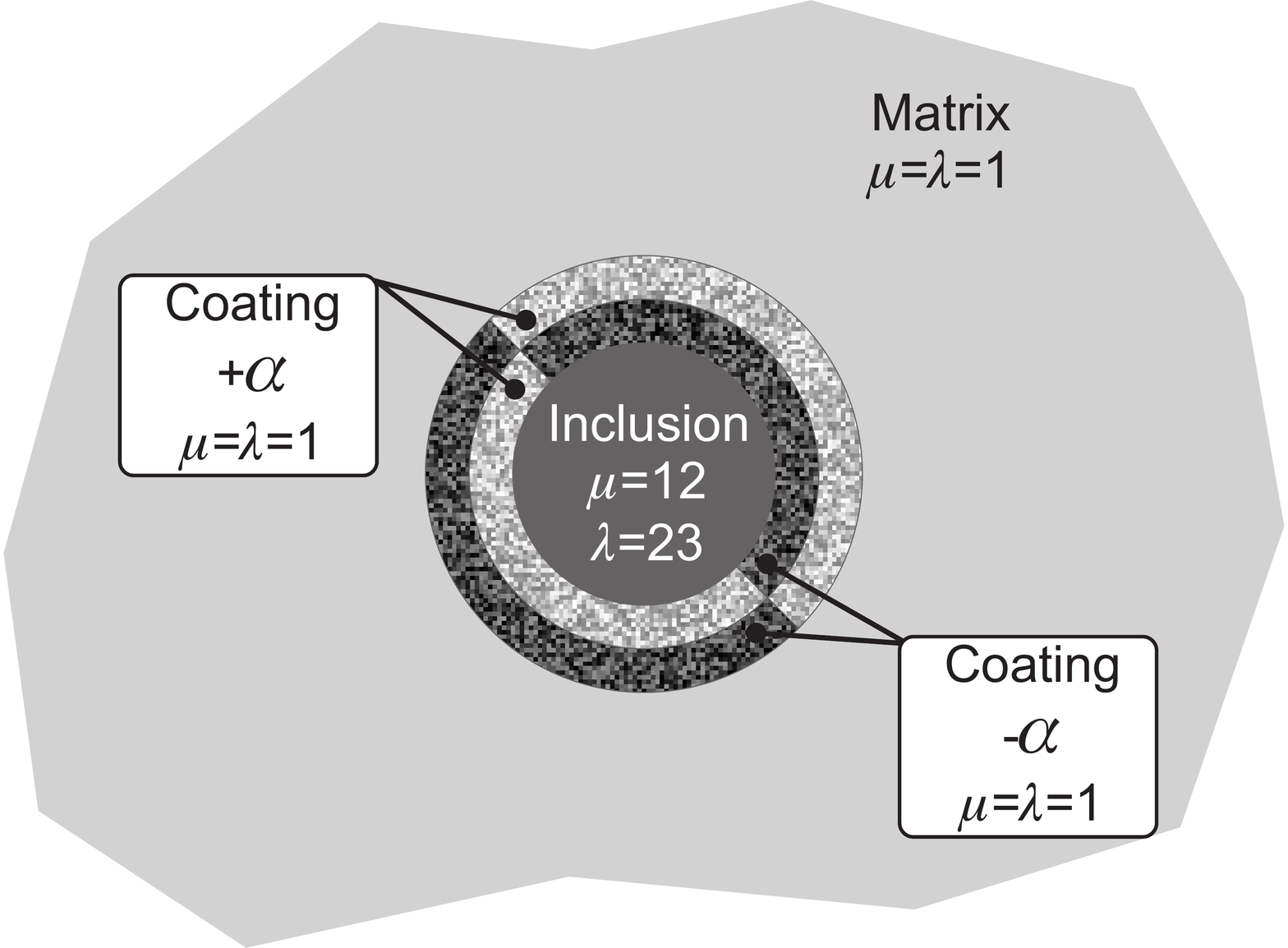}}
\caption{Coated inclusion. The coating consists of the chiral material with the spinner constants $\alpha$ or $-\alpha$, as illustrated on the diagram. }
\label{model}
\end{figure}

\subsection{An elastic inclusion with a chiral coating}

Here we consider an elastic plane containing a coated inclusion.
The two-phase coating of the inclusion is assumed to have the chiral term in the governing equations, whereas the
ambient medium, around the coating and the inclusion itself are assumed to be non-chiral.
The coating consists of two semi-rings, with the spinner constants being of the same magnitude and opposite sign, as shown in Fig. \ref{model}.
The coated inclusion is chosen to be symmetric about the straight line passing through the centre of the inclusion and the point where the concentrated force is applied.

\begin{figure}[ht!]
\centering
{\includegraphics[width=4.1cm,angle=0]{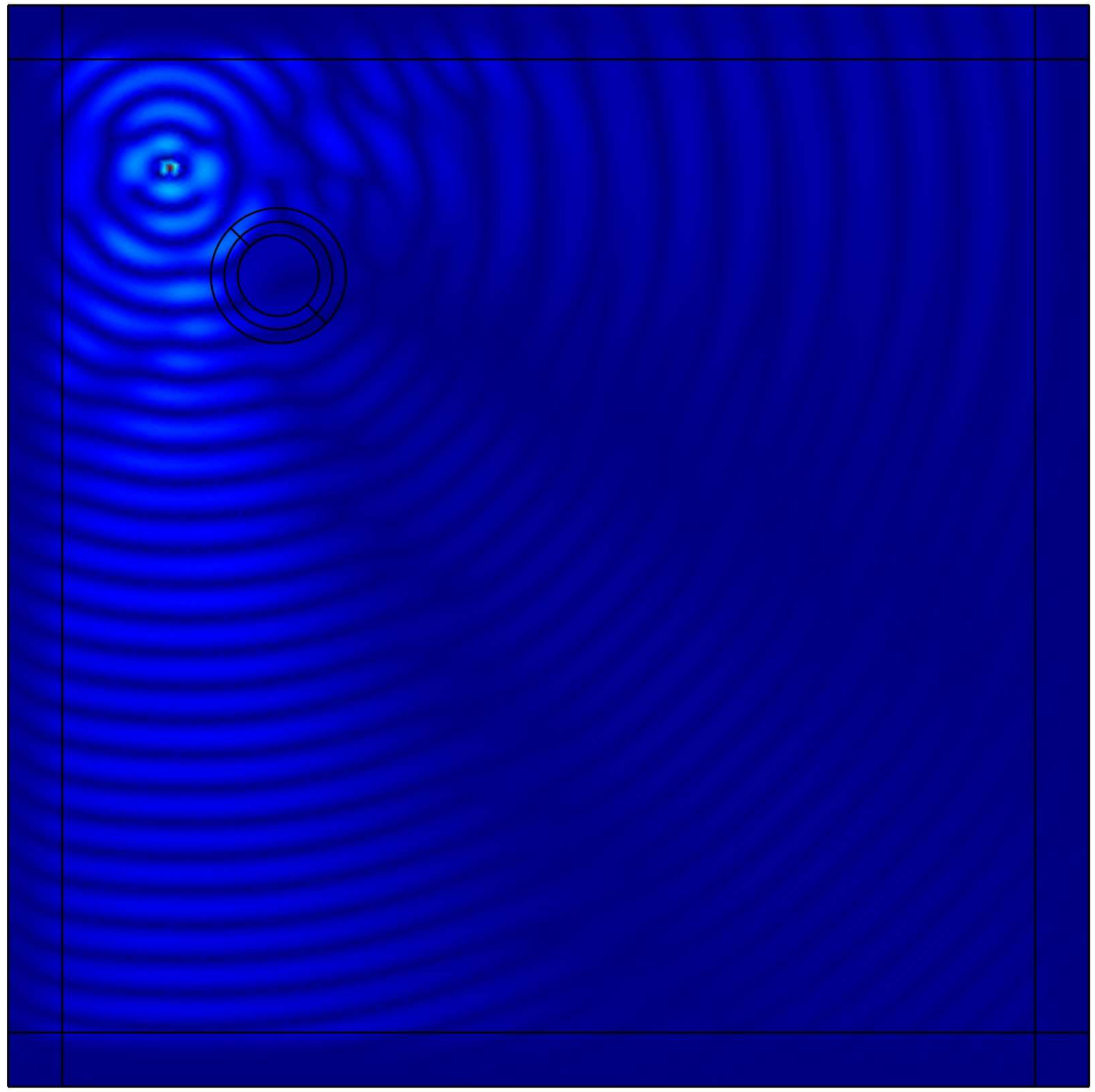}}   {\includegraphics[width=4.1cm,angle=0]{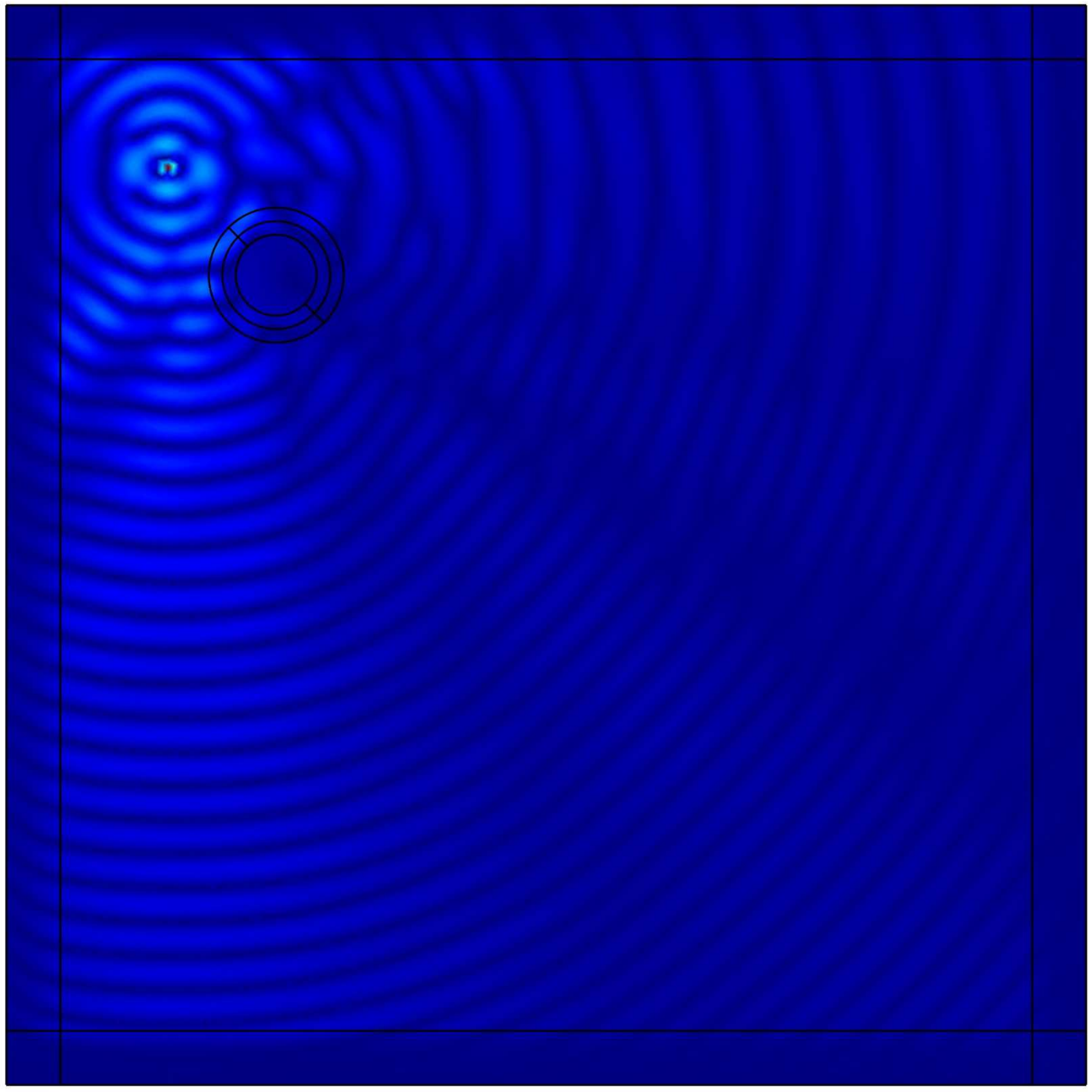}}  {\includegraphics[width=4.1cm,angle=0]{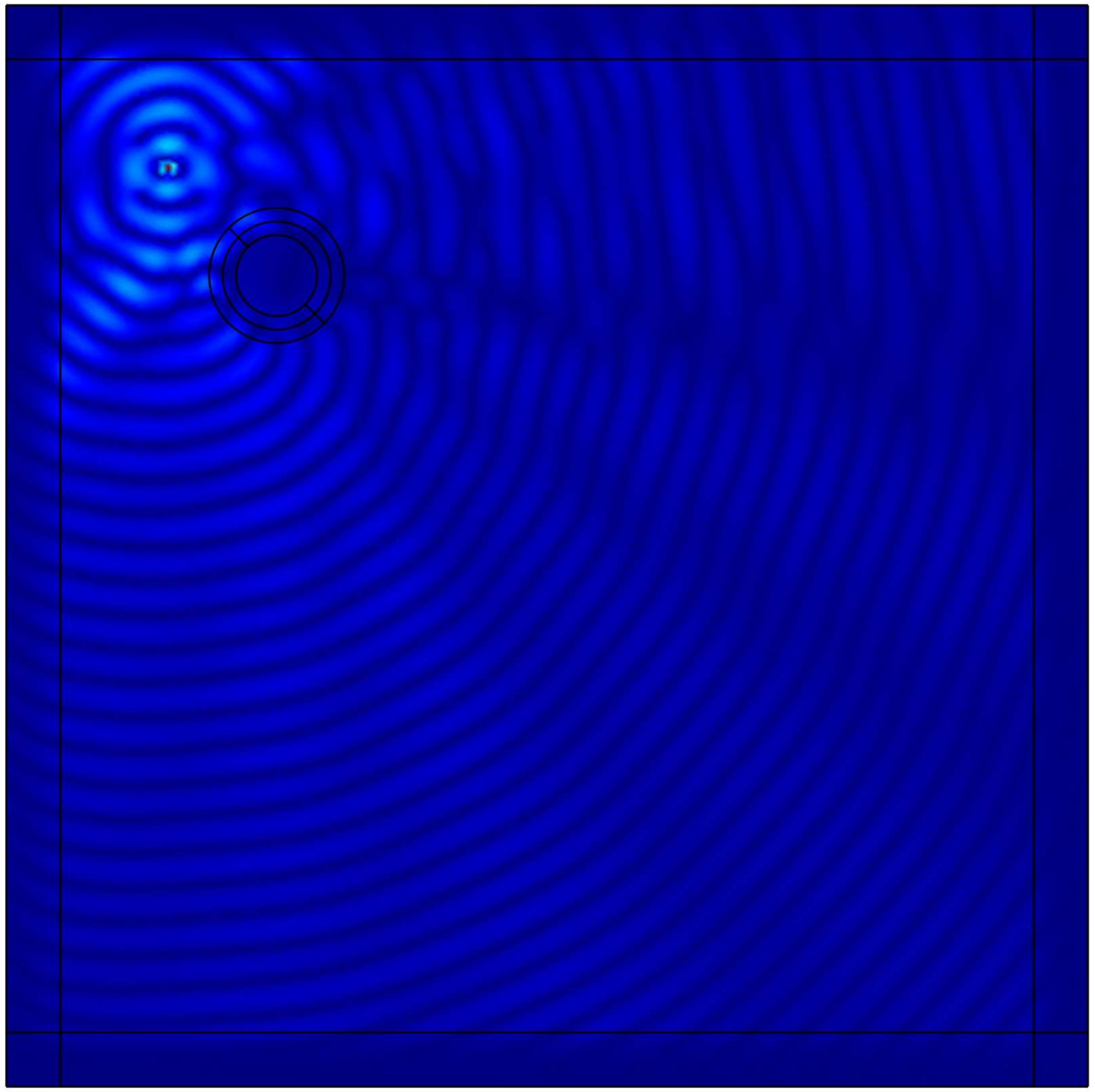}
}
\centerline{(a)~~~~~~~~~~~~~~~~~~~~~~~~~~~(b)~~~~~~~~~~~~~~~~~~~~~~~~~~~~(c)  }
\caption{Displacement amplitude in the plane containing a coated inclusion (as shown in Fig. \ref{model}). The loading is provided by the horizontal point force. The inclusion is placed in the region dominated by the shear waves. Three computations correspond to different values of the spinner constant $\alpha$: (a) $\alpha=0.0$; (b) $\alpha = 0.9$; (c) $\alpha = 2.0$.
}
\label{force_in1}
\end{figure}

An incident wave is generated by a point source placed at a finite distance from the inclusion. For the computational examples, the source term is provided by a 
point force or concentrated moment with a time harmonic amplitude of radian frequency
$\Go = 50$.
In the computation, the elastic inclusion has the normalised Lam\'e constants $\Gl= 23, \mu= 12$ and the same mass density as in  the ambient medium. The coating is assumed to have the same elastic moduli and the same mass density as in the ambient elastic matrix.

\begin{figure}[ht!]
\centering
{\includegraphics[width=6cm,angle=0]{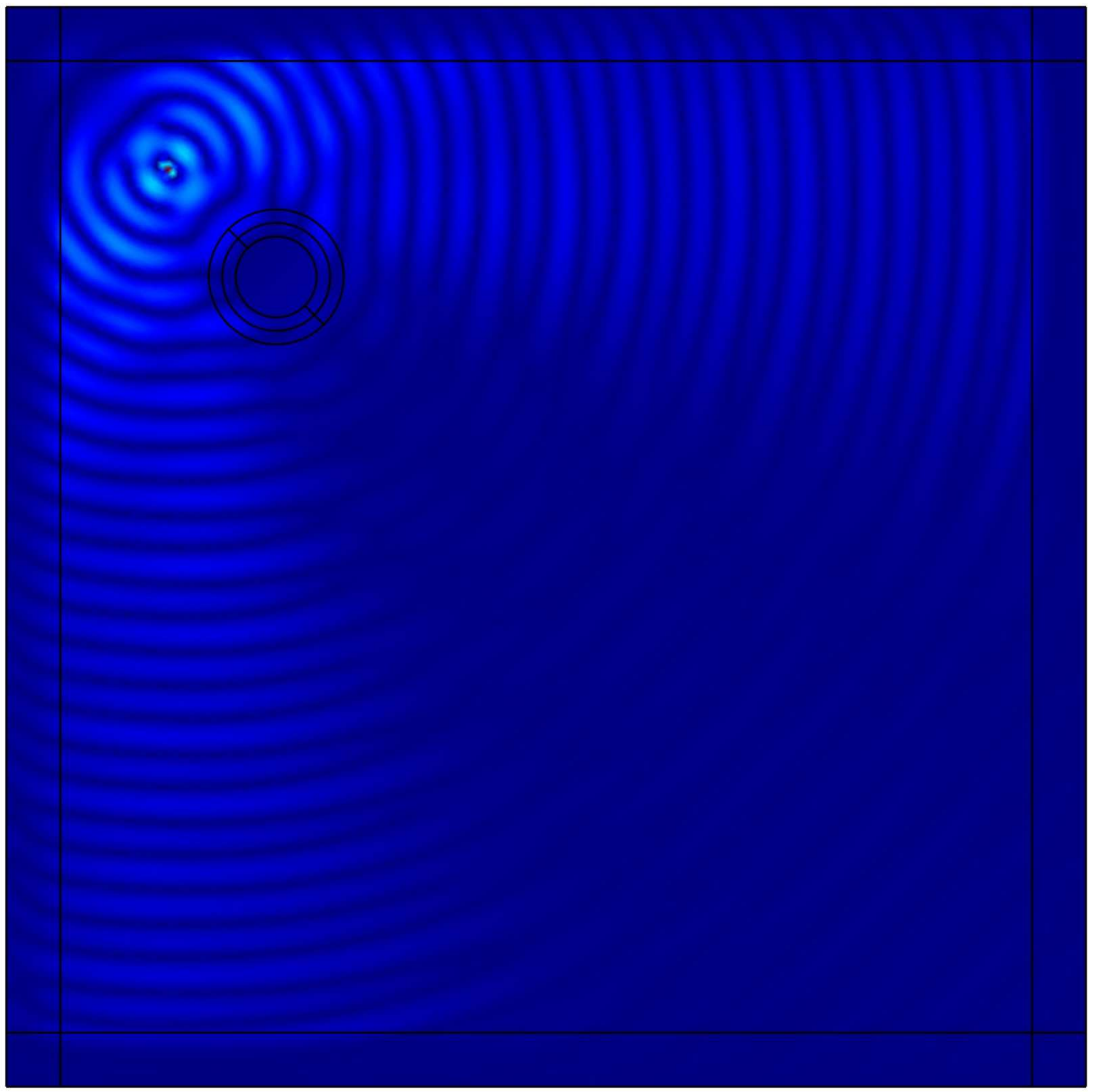}}   {\includegraphics[width=6cm,angle=0]{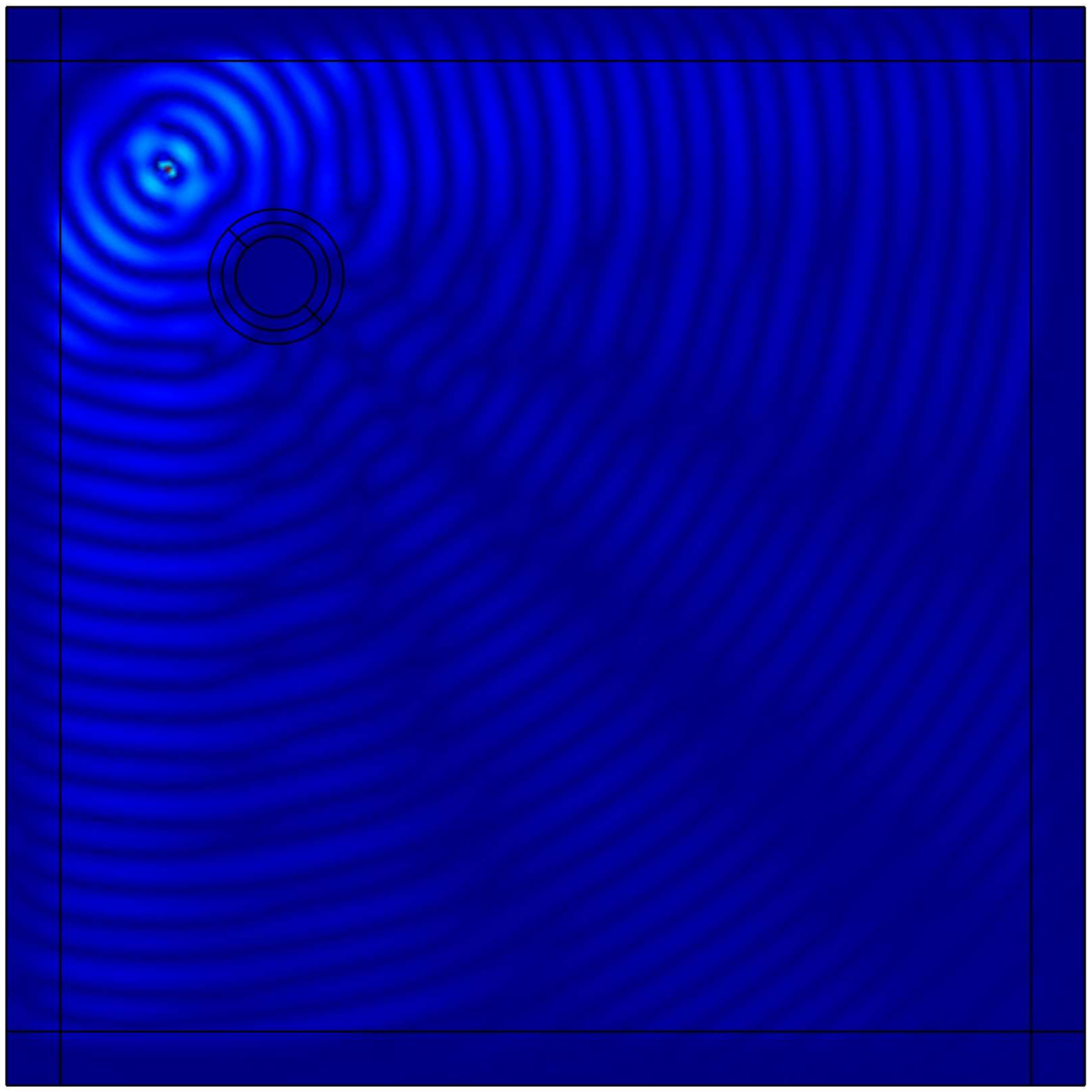}}
\centerline{(a)~~~~~~~~~~~~~~~~~~~~~~~~~~~~~~~~~~~~~~~~~~~~(b)}
\caption{Comparison of the cases  (a) $(\Ga=0.0)$ of the elastic inclusion without coating and (b) the inclusion with the chiral coating $(\Ga=1.5)$,  placed in the elastic medium loaded by the time-harmonic point force pointed towards the centre of the inclusion. The chiral coating reduces the
shaded
region created by the inclusion.
}
\label{force_p_0_15}
\end{figure}

For the  isotropic medium  with an inclusion without a coating ($\Ga = 0.0$) the results of the computation are shown in Fig. \ref{force_in1}a. Comparing with Fig. \ref{force1} one can clearly see the shaded region associated with the perturbation field produced by the elastic inclusion.

\begin{figure}[ht!]
\centering
{\includegraphics[width=6cm,angle=0]{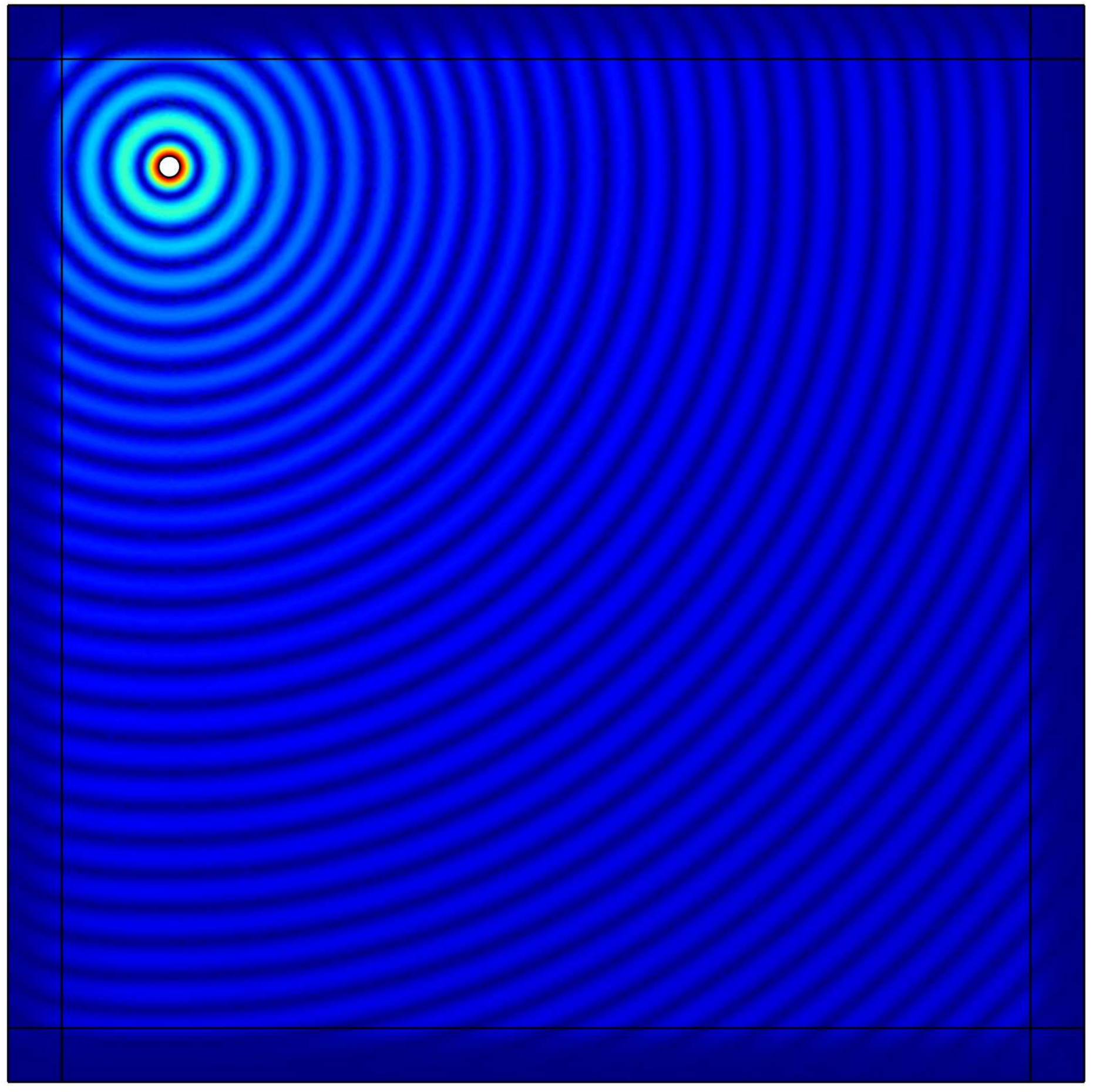}}
{\includegraphics[width=6.cm,angle=0]{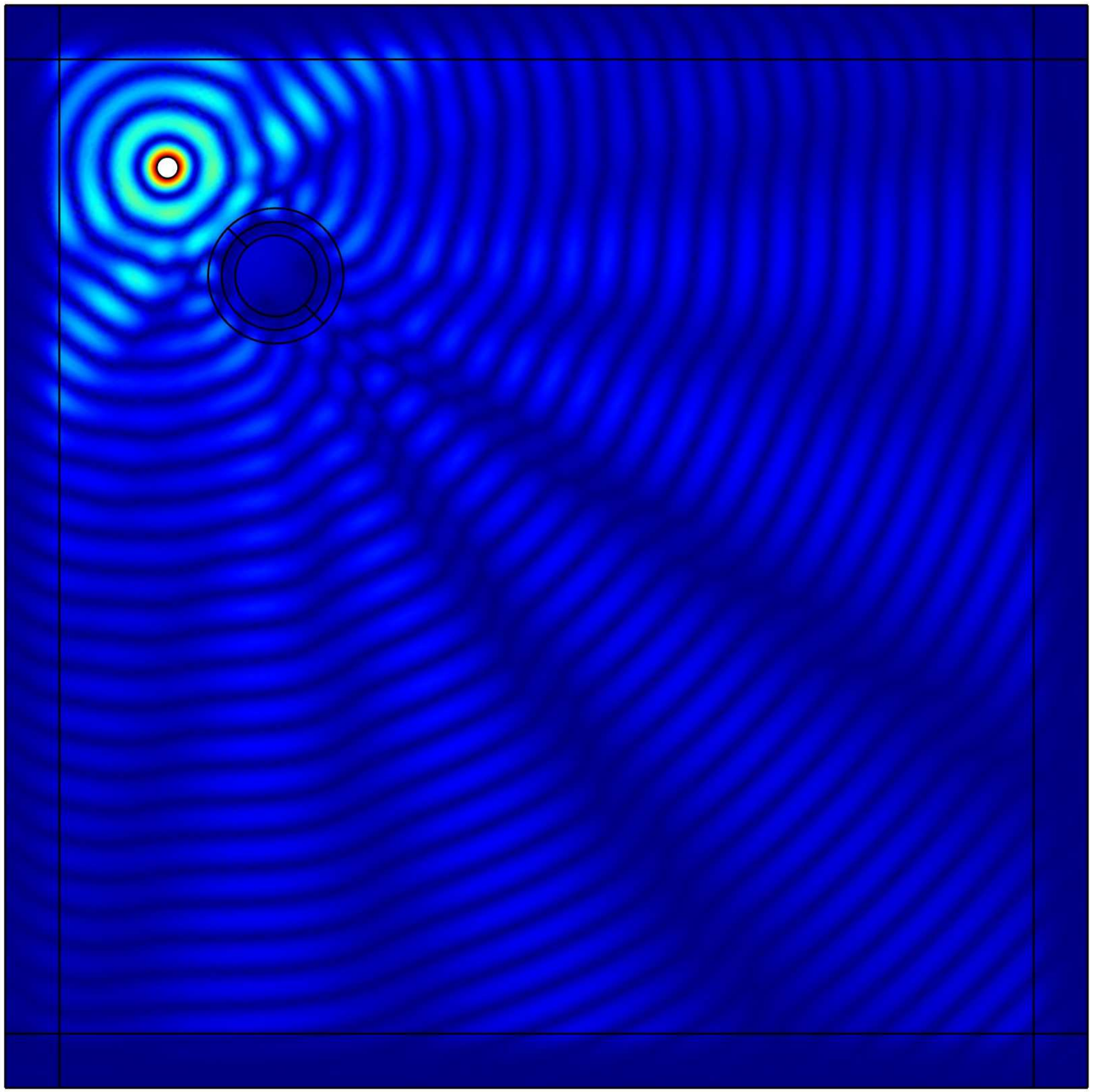}}

{\includegraphics[width=6.cm,angle=0]{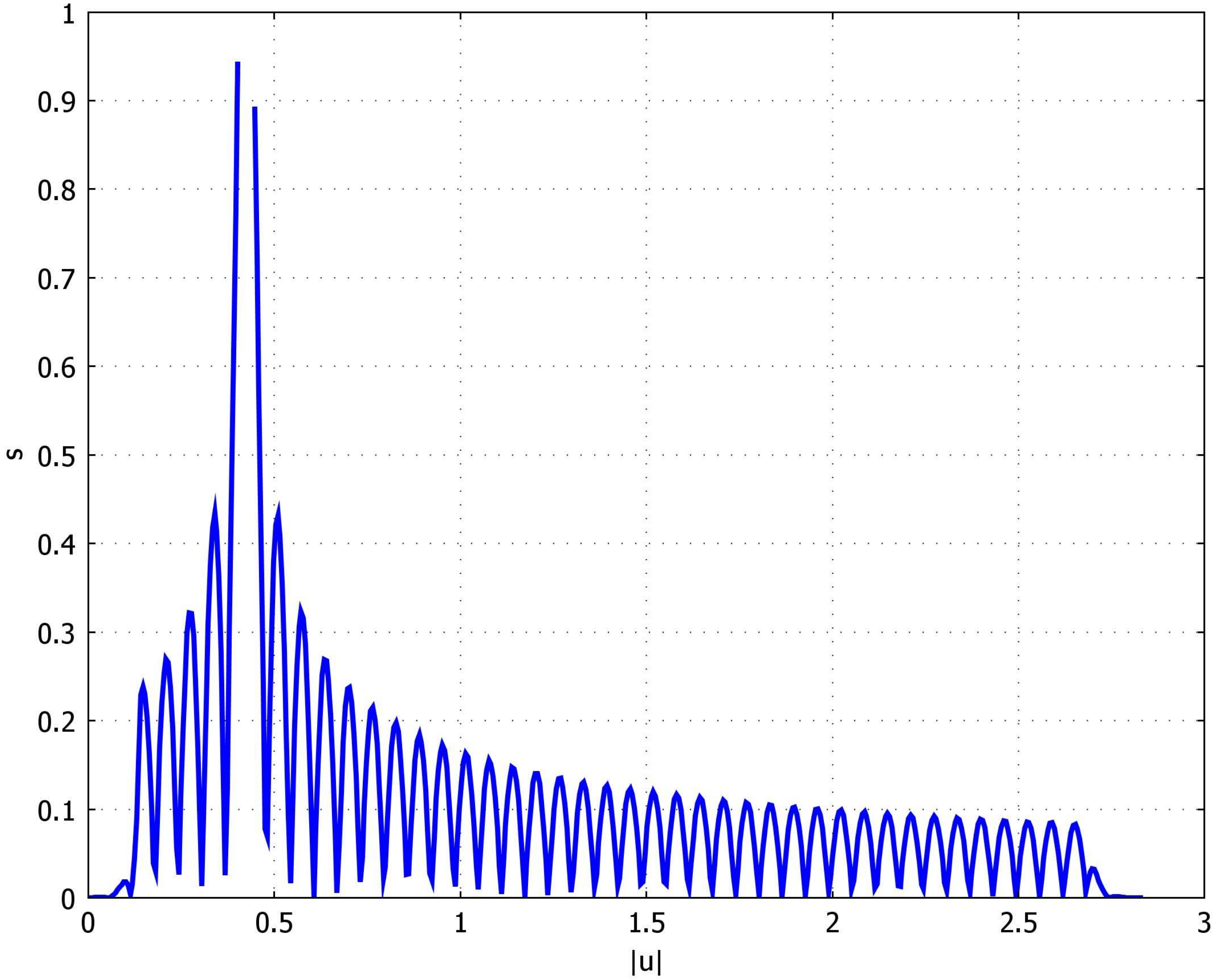}}
{\includegraphics[width=6.cm,angle=0]{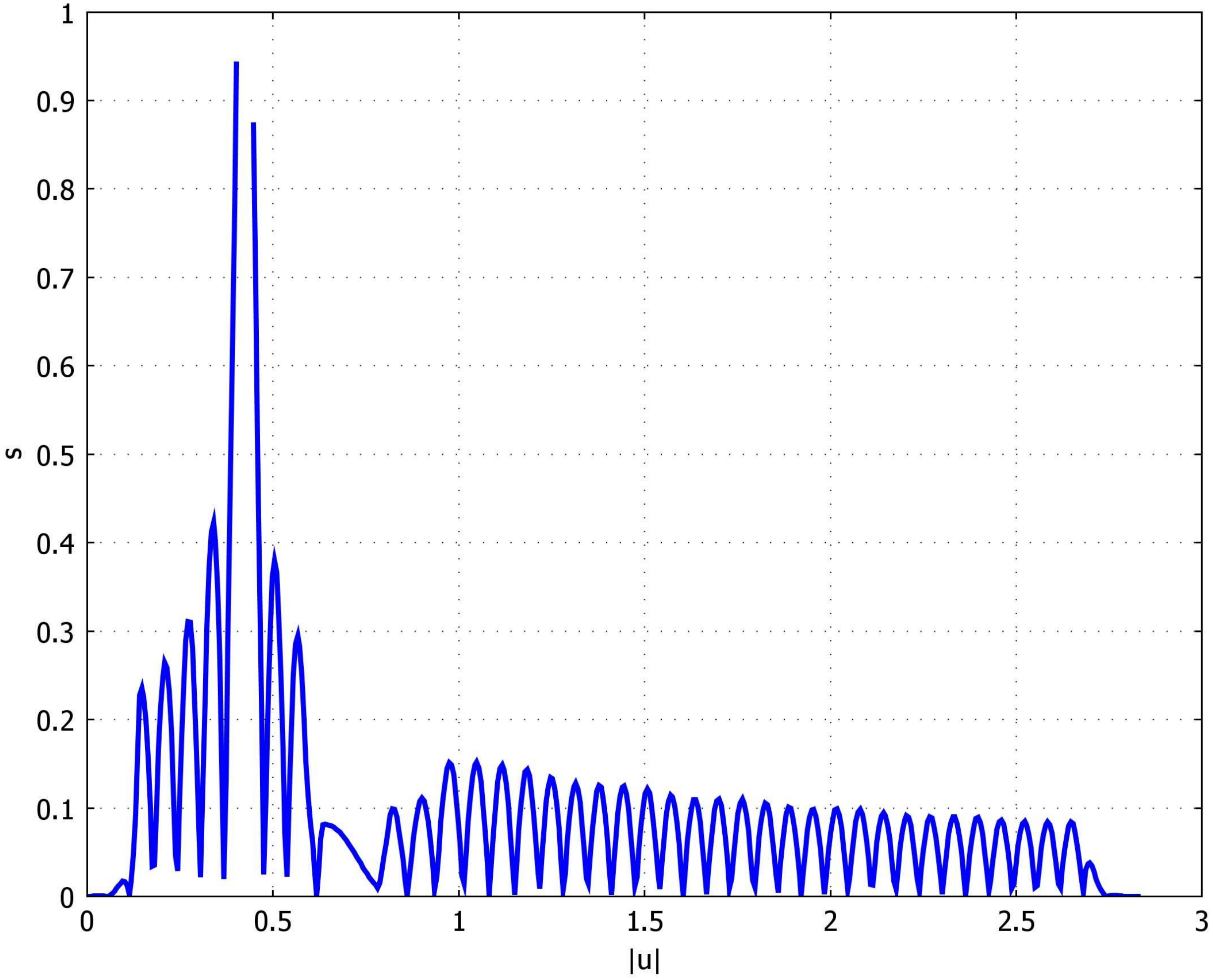}}
\centerline{(a)~~~~~~~~~~~~~~~~~~~~~~~~~~~~~~~~~(b) }
\caption{Comparison of the cases (a) homogenous medium with no inclusion and (b) the inclusion with the chiral coating $(\Ga=2.0)$,
placed in the elastic medium loaded by the time-harmonic point moment.
}
\label{force_shear_2}
\end{figure}

The next set of computations in Figs. \ref{force_in1}b and \ref{force_in1}c  corresponds to the chiral two-phase coating placed around the inclusion
and can be interpreted as a ``cloak'' guiding shear waves around the inclusion.
It is noted that, in these two diagrams where the point force is horizontal, the inclusion is placed in the region dominated by shear waves. It is also seen that the effect of 
cloaking is most pronounced in the supercritical regime when the spinner constant of the inclusion is chosen so that $|\Ga| \geq 1$.

One of the important features of the computations shown in Fig. \ref{force_in1}
is the directional preference of the ambient field produced by the point force. In particular, the above diagrams correspond to the case when the ambient field in the immediate neighbourhood of the inclusion is dominated by shear waves. A change of the orientation of the force with respect to the position vector of the inclusion results in a different pattern of the ambient field surrounding the inclusion. Fig. \ref{force_p_0_15} correspond to the situation when the force is pointed towards (or opposite) the centre of the elastic coated inclusion. As before we show the results of the computations for the inclusion without the coating ($\Ga=0$) in Fig. \ref{force_p_0_15}a, and supercritical chiral coating ($\Ga=1.5$) in Fig. \ref{force_p_0_15}b.
The direct comparison of these diagrams shows that the chiral coating reduces the shadow region associated with the inclusion. Furthermore we observe that the ripples behind the inclusion have a size corresponding to that of the wave length of shear waves, which suggests that the chiral coating is acting as
a shear polarising cloak.

Finally, we consider the configuration when the ambient elastic field has no directional preference. The loading is provided by a time-harmonic concentrated moment of the radian frequency
$\Go = 50$.
The computation for the homogeneous medium, without an  inclusion
is presented in Fig. \ref{force_shear_2}a. We also note the earlier Fig.  \ref{force_shear_1}a where the computation is presented for an uncoated inclusion. The shadow region can be reduced via introduction of the chiral coating. The
supercritical case of $\Ga= 2.0$ is included in Fig. \ref{force_shear_2}b, which shows a new diffraction pattern, with conical regions where waves have enlarged amplitude.
Two diagrams in the bottom row of Fig. \ref{force_shear_2} give the amplitude of the total displacement along the main diagonal of the square computational region.
They show that  along this line the displacement amplitude behind the inclusion  with the chiral coating of $\Ga = 2.0$ is similar to the displacement amplitude in the homogeneous medium without any inclusion.

\section{Concluding remarks}

This paper has placed together two concepts: the model of a discrete lattice with spinners and the homogenisation approximation by a continuum chiral medium.
The vorticity constants in the governing equations describing the chiral media are evaluated explicitly and, furthermore, dynamic response of both discrete and continuous systems have been analysed in detail.

Novel properties of Bloch waves in discrete vortex-type elastic lattices have been identified, with analytical findings being complemented by the numerical illustrations of dispersion surfaces.

In the homogenisation approximation, the chiral material has been used to design a composite cloak around an inclusion, so that the incident waves are guided around the inclusion. The result of such an interaction can lead to a substantial reduction of the shadowed region behind the inclusion, and the observation that the coating can be interpreted as a polarising cloak.

Furthermore, it was demonstrated that the parameters of the chiral coating can also be  tuned so that the shadowed region is enhanced and the amplitude of the displacement behind the inclusion becomes negligibly small.

We envisage a range of applications of the proposed model in problems of geophysics and structural design of cloaks shielding defects, which interact with elastic waves.

\vspace*{5mm}
\noindent
{\sl Acknowledgments}

{\footnotesize \noindent
The financial support of Research Centre in Mathematics and
Modelling of the University of Liverpool and Italian Ministry of
University and Research under PRIN 2008 "Complex materials and
structural models in advanced problems of engineering" (M.B.) are
gratefully acknowledged.}

\end{document}